\documentclass[sn-mathphys-num]{sn-jnl}% Math and Physical Sciences Numbered Reference Style 
\usepackage{graphicx}%
\usepackage{multirow}%
\usepackage{amsmath,amssymb,amsfonts,bm}%
\usepackage{amsthm}%
\usepackage{mathrsfs}%
\usepackage[title]{appendix}%
\usepackage{float}
\usepackage{epsfig}
\usepackage[caption=false]{subfig}
\usepackage{xcolor}%
\usepackage{textcomp}%
\usepackage{manyfoot}%
\usepackage{booktabs}%
\usepackage{algorithm}%
\usepackage{algorithmicx}%
\usepackage{algpseudocode}%
\usepackage{listings}%
\usepackage{moresize}
\theoremstyle{thmstyleone}%
\theoremstyle{thmstyletwo}%
\theoremstyle{thmstylethree}%
\newtheorem{res}{Result}



\begin{document}

\title[Article Title]{Robust Bayesian approach for reliability prognosis of nondestructive one-shot devices under cumulative risk model}

\author[1]{\fnm{Shanya} \sur{Baghel}}\email{shanyabaghel.20dr0130@mc.iitism.ac.in}

\author*[1]{\fnm{Shuvashree} \sur{Mondal}}\email{shuvasri29@iitism.ac.in}

\affil*[1]{\orgdiv{Department of Mathematics and Computing}, \orgname{Indian Institute of Technology (ISM) Dhanbad}, \postcode{826004}, \state{Jharkhand}, \country{India}}


\abstract{The present study aims to determine the lifetime prognosis of highly durable nondestructive one-shot devices (NOSD) units under a step-stress accelerated life testing (SSALT) experiment applying a cumulative risk model (CRM).  In an SSALT experiment, CRM retains the continuity of hazard function by allowing the lag period before the effects of stress change emerge.  When prior information about the model parameters is available, Bayesian inference is crucial.  In a Bayesian analysis of such lifetime data, conventional likelihood-based Bayesian estimation frequently fails in the presence of outliers
in the dataset.  This work incorporates a robust Bayesian approach utilizing a robustified posterior based on the density power divergence measure.  The order restriction on shape parameters has been incorporated as a prior assumption to reflect the decreasing expected lifetime with increasing stress levels.  In testing of hypothesis, a Bayes factor is implemented based on the robustified posterior.  In Bayesian estimation, we exploit Hamiltonian Monte Carlo, which has certain advantages over the conventional Metropolis-Hastings algorithms.  Further, the influence functions are examined to evaluate the robust behaviour of the estimators and the Bayes factor.  Finally, the analytical development is validated through a simulation study and a real data analysis.}

\keywords{ Bayes factor, cumulative risk model, Hamiltonian Monte Carlo, influence function, nondestructive one-shot device, robust Bayes estimation }



\maketitle

\section{Introduction}\label{sec1}
In recent times, the reliability prognosis of nondestructive one-shot devices (NOSD) is drawing an increasing amount of attention because of their broad applicability in industrial and engineering domains.  Metal fatigue, spare wheels and light bulbs are some examples of NOSD.  Unlike one-shot devices, NOSD may survive multiple tests, offering additional data for reliability estimation.  The observation for such devices is mostly restricted to recording if device failure occurs before or after a specified inspection time, which leads to the study of dichotomous data only.  For highly durable products, accelerated life testing (ALT) experiments are frequently employed to get more failures within a short span of time \cite{prajapati2023misspecification,fu2025research,valivs2025light,wang2025review}. 

The reliability analysis of NOSD under step-stress accelerated life test (SSALT) has garnered the attention of various studies \cite{balakrishnan2022non,balakrishnan2022restricted,balakrishnan2023robust,balakrishnan2024non} recently.  In SSALT, stress increases stepwise over predefined time points, and a connection model is needed to relate lifetime distributions at different levels.  In the literature, we see a broad application of the cumulative exposure model (CEM) in robust estimation, e.g. \cite{ling2019optimal,ling2020optimal,balakrishnan2023robust,balakrishnan2024non,balakrishnan2024step}.  However, a change in the stress level in this model is instantaneous, leading to a discontinuity in hazard function at the stress change point.  To address this shortcoming, Van Dorp and Mazzuchi \cite{RENEVANDORP200455} proposed a model based on the hazard rate function, which was later referred to as the cumulative risk model (CRM) by Kannan et al.\cite{kannan2010survival}.  This model removes discontinuity by allowing the lag period before the effects of stress change emerge.  Although various authors have studied CRM in the past \cite{yao2013step,beltrami2017weibull,qiao2023inference}, the SSALT experiment exploiting CRM for NOSD is yet to be explored.

The estimation procedure based on the classical approach is generally satisfactory.  However, with the availability of prior knowledge, the Bayesian approach comes into the picture \cite{mun2019bayesian,ariyo2022bayesian,abdel2023bayesian,allotey2023bayesian,sanju2024evaluating,ling2024efficient,salem2024bayesian,kumari2024bayes}.  There is sufficient literature on Bayesian analysis of one-shot devices; readers may refer to \cite{quigley2009empirical,lee2014study,balakrishnan2015bayesian,sharma2018hierarchical,sharma2021hierarchical,ashkamini2023bayes,rougier2024bayesian,salah2025point} and references therein.  The conventional likelihood-based Bayesian estimation may not provide desired statistical inference with small deviations from the assumed model conditions, which raises the need for a robust Bayesian method.  Ghosh and Basu \cite{g2016} divulged into the development of robust Bayesian inference where the density power divergence (DPD) measure \cite{basu1998robust,patra2013power,hazra2024robust} has substituted likelihood in the posterior density function.  To the best of our knowledge, a robust Bayesian approach in the context of NOSD under SSALT is barely applied, which brings novelty to this study.

The SSALT experiment accelerates the failure of the units by increasing the stress level \cite{gorny2020exact,bedbur2022testing}.  Therefore, it is pretty reasonable to assume that the expected lifetime of the experimental units is lower at the higher stress level.  This information can be incorporated to develop an order-restricted prior assumption.  This study assumes Normal and Dirichlet distributions \cite{fan2009bayesian} as priors based on data.  Further, an ordered Dirichlet-Gamma distribution is assumed on the shape parameters to incorporate a larger hazard rate with increasing stress.  Although several studies \cite{samanta2018order,mondal2019point,mondal2020bayesian,pal2021bayesian,wiedner2021bayesian} have assumed order restriction on scale parameters, order restricted prior assumption in robust Bayesian analysis is unprecedented. 

 Under the prior assumptions for the considered model, a closed form of posterior cannot be obtained.  While Gibbs sampler and Metropolis-Hastings algorithms are frequently used for posterior estimation, they may be inefficient in exploring the target distribution with high dimensional or highly correlated variables \cite{thach2019reparameterized}.  Hamiltonian Monte Carlo (HMC), introduced by Neal \cite{neal2011mcmc,neal2012bayesian} to the application of statistics, offers a solution, providing accurate results and flexibility in complex models \cite{thanh2021additive,abba2023new}.  For an in-depth explanation of HMC, one can refer to \cite{monnahan2017faster,abba2024robust} and the references therein.  The present study is the first attempt to seek HMC to solve the robust Bayes estimation problem of NOSD test data under SSALT.  

 Another critical aspect of the Bayes framework is the testing of hypotheses through the Bayes factor, which was initially introduced by Jeffreys \cite{jeffreys1973scientific, jeffreys1935some, jeffreys1998theory} and later applied by numerous researchers.  In this study, we develop a robust Bayes factor by applying the proposed robustified posterior.  Further, influence function analysis is evident in the study of robustness.  Basu et al. \cite{basu1998robust} and Ghosh and Basu \cite{g2016} derived influence functions for DPD-based and robust Bayes estimates, respectively.  However, the influence function analysis of the robust Bayes factor evaded the attention of researchers, and its application for NOSD test data has yet to be conducted.

 The present study focuses on the Bayesian inference of NOSD under CRM SSALT experiment with interval monitoring over intermediate inspection time points.  The lifetime of NOSD is assumed to follow the well-known standard family of Lehman distributions \cite{bobotas2015step,kateri2015inference,pal2021simple}.  The estimation procedure relies on a robust Bayes estimation (RBE)\cite{g2016} method, creating a robustified posterior density through the exponential form of the maximizer equation using the DPD measure.  In the prior selection, Normal and Dirichlet prior are considered.  The order-restricted prior assumption on scale parameters has also been developed through Dirichlet-Gamma distribution.  Additionally, this study explores the testing of the hypothesis utilizing a robust Bayes factor derived from the robustified posterior.  Furthermore, the influence functions are derived and examined thoroughly to assess the robust behaviour of the point estimators in the Bayesian framework.  In the testing of hypotheses, the influence function reflects how outlying observation can influence the Bayes factor under the null hypothesis, potentially affecting decision-making.

 The rest of the article proceeds as follows.  Section \ref{sec2} focuses on building a cumulative risk model.  The robust Bayesian estimation method is discussed in Section \ref{sec3}.  In Section \ref{sec4}, testing of hypothesis based on robust Bayes factor is developed.  Section \ref{sec5} studies the property of robustness through the influence function.  Finally, Sections \ref{sec6} and \ref{sec7}  contain the simulation study and data analysis, respectively.  Concluding remarks are given in the Section \ref{sec8}.

\section{Cumulative risk step-stress model} \label{sec2}
This section discusses the experimental setup for analyzing the nondestructive one-shot device (NOSD) testing data under the cumulative risk step-stress model.

\subsection{\textbf{The model}}
Consider $n$ NOSD units are put to step-stress accelerated life testing (SSALT) experiment with $k$ stress levels denoted by $x_i\,;\,i=1,2,\dots,k.$  Starting from initial stress $x_1$, the stress level of NOSD units is increased from $x_i$ to $x_{i+1}$ at pre-specified time points $\tau_i\;;\;i=1,2,\dots,k$ and $\tau_0=0.$
At each stress level, the lifetime distribution of the NOSD is assumed to follow the Lehman family of distributions with different shape and scale parameters whose cumulative distribution function (cdf) and probability density function (pdf) are defined as
\begin{equation}
  \begin{aligned}
    F_i(t)&=1-\exp\big\{-\lambda_i Q(t;\gamma_i)\big\}.\\
    f_i(t)&=\lambda_i Q^{'}(t;\gamma_i)\exp\big\{-\lambda_i Q(t;\gamma_i)\big\}.
\end{aligned}  \label{dist}
\end{equation}
where $\lambda_i>0$ and $\gamma_i>0$ and $Q(t;\gamma_i)$ is strictly increasing function of $t$.  Various lifetime distributions such as exponential, Weibull and Gompertz are members of this family for $Q(t;\gamma_i)=t, t^{\gamma_i}, (e^{\gamma_it}-1)$ respectively.  Here, in equation \eqref{dist}, parameter $\lambda_i$ is related to stress factor in log-linear form as
\begin{equation}
    \lambda_i=\exp (c_0+c_1x_i)\;;\;i=1,2,\dots,k.
\end{equation}
Thus the set of model parameters to be estimated is denoted by $$\bm{\theta}=\{(c_j,\gamma_i)^{T}\,;\,j=0,1\,;\,i=1,2,\dots,k\}.$$

\subsection{\textbf{Cumulative risk model}}
Introduced by Van Dorp and Mazzuchi \cite{RENEVANDORP200455}, formalized by Kannan et al.\cite{kannan2010survival} and generalized by Kannan and Kundu \cite{kannan2020simple}, Cumulative risk model (CRM) addresses discontinuities in hazard rates occurring due to instantaneous stress changes.  CRM ensures a smooth transition by inducting lag period $\delta$ before the effect of stress change emerges.  The hazard rate function is usually assumed to be linearly modelled in this lag period. 

\subsubsection{Lag period $\delta$ known}
\begin{figure}[htb!]
\begin{center}
\subfloat[Step-stress\label{st}]{\includegraphics[height=4.5cm,width =0.53\textwidth]{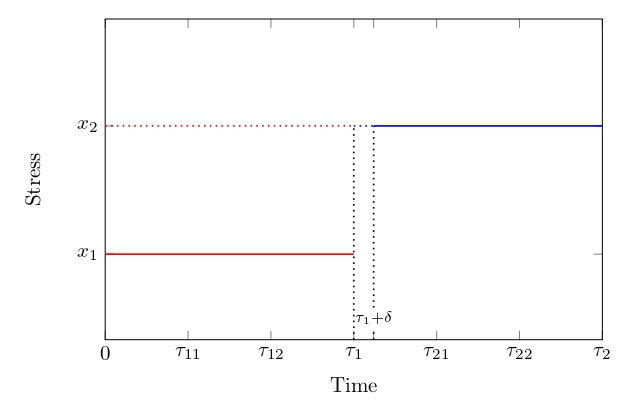}} 
\subfloat[Hazard rate\label{haz}]{\includegraphics[height=4.9cm,width =0.46\textwidth]{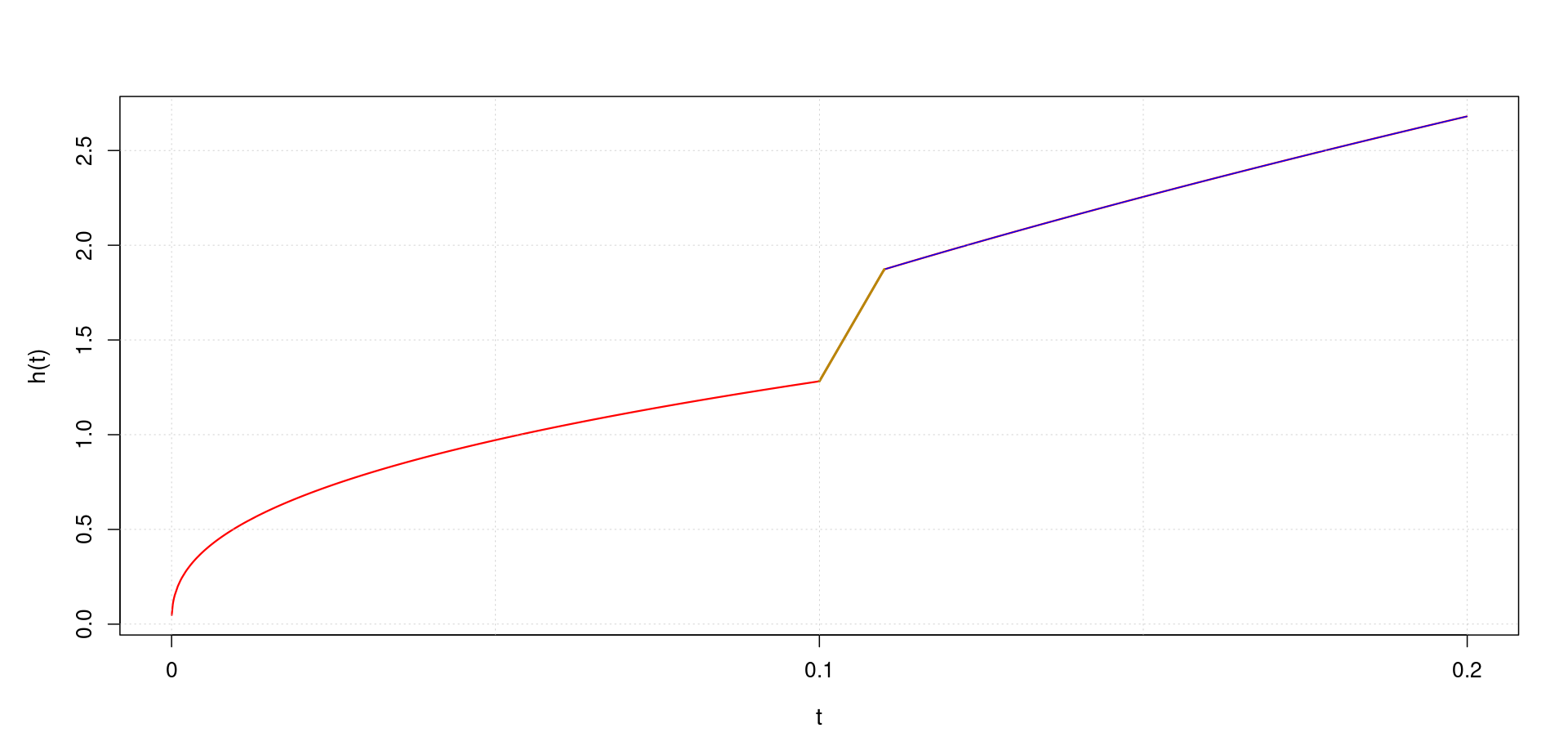}}
\end{center}
\caption{Step-stress (left) and Hazard rate (right) for Weibull lifetime distribution under CRM.}
\label{hazst}
\end{figure}

In the development of CRM, we assume the lag period $\delta$ is known.  Figure \ref{hazst} depicts a step-stress environment and linearly modelled hazard rate explaining Weibull lifetime distribution defined by CRM under a simple SSALT.
For NOSD testing data, the piece-wise hazard rate function under the Lehman family of distribution takes the following form.
 \begin{equation}
h(t)=
\begin{cases}
\lambda_1 Q^{'}(t;\gamma_1) &; \;0<t\leq \tau_1.\\
a_{i-1}+b_{i-1}t &;\; \tau_{i-1}<t\leq \tau_{i-1}+\delta\;,\;i=2,3,\dots,k-1.\\
\lambda_i Q^{'}(t;\gamma_i) &;\; \tau_{i-1}+\delta<t\leq \tau_i\;;\;i=2,3,\dots,k-1.\\
\lambda_k Q^{'}(t;\gamma_k)&;\;\tau_{k-1}<t<\infty.
\end{cases}\label{hazeq}
\end{equation}
To ensure continuity of $h(t)$ in equation \eqref{hazeq}, $a_{i-1}$ and $b_{i-1}$ must satisfy 
\begin{equation}
\begin{aligned}
    a_{i-1}+b_{i-1}\tau_{i-1}&=\lambda_{i-1} Q^{'}(\tau_{i-1};\gamma_{i-1}).\\
      a_{i-1}+b_{i-1}(\tau_{i-1}+\delta)&= \lambda_i Q^{'}(\tau_{i-1}+\delta;\gamma_{i-1}).
\end{aligned}\label{hazsat}
\end{equation}
By solving equation \eqref{hazsat}, we obtain
\begin{align*}
    a_{i-1}&=\frac{1}{\delta}\bigg\{(\delta+\tau_{i-1})\lambda_{i-1}Q^{'}(\tau_{i-1};\gamma_{i-1})-\tau_{i-1}\lambda_i Q^{'}(\tau_{i-1}+\delta;\gamma_i)\bigg\}.\\
    b_{i-1}&=\frac{1}{\delta}\bigg\{\lambda_i Q^{'}(\tau_{i-1}+\delta;\gamma_i)-\lambda_{i-1} Q^{'}(\tau_{i-1};\gamma_{i-1})\bigg\}.
\end{align*}
Therefore, Survival function $S(t)=e^{-\int_{0}^{t}h(x)dx}$ is obtained as
\begin{equation}
S(t)=
\begin{cases}
\exp\Big\{-\lambda_1 Q(t;\gamma_1)\Big\} \quad ; \;0<t\leq \tau_1.\\
\exp\Big\{-D^{(\delta)}(t;\gamma_{i-1,i})\Big\}\exp\bigg[-\Big\{\lambda_{i-1}Q(\tau_{i-1};\gamma_{i-1})+\sum\limits_{l=1}^{i-2}E^{(\delta)}(\tau_l;\gamma_{l+1,l})\Big\}\bigg]\\
\hspace{3.9cm};\; \tau_{i-1}<t\leq \tau_{i-1}+\delta\;;\;i=2,3,\dots,k-1.\\
\exp\Big\{-\lambda_i Q(t;\gamma_i)\Big\}\exp\bigg\{-\sum\limits_{l=1}^{i-1}E^{(\delta)}(\tau_l;\gamma_{l+1,l})\bigg\} \\
\hspace{3.9cm};\; \tau_{i-1}+\delta<t\leq \tau_i\;;\;i=2,3,\dots,k-1.\\
\exp\Big\{-\lambda_k Q(t;\gamma_k)\Big\}\exp\bigg[-\Big\{\lambda_{k-1} Q(\tau_{k-1};\gamma_{k-1})+\sum\limits_{i=1}^{k-2}E^{(\delta)}(\tau_i;\gamma_{i+1,i})\Big\}\bigg]\\
\hspace{3.9cm};\;\tau_{k-1}<t<\infty.
\end{cases},\label{sirv}
\end{equation}
where,
\begin{align*}
    D^{(\delta)}(t;\gamma_{i-1,i})&=\frac{(t-\tau_{i-1})^2}{2\delta}\Bigg[\Big\{2\delta(t-\tau_{i-1})^{-1}-1\Big\}\lambda_{i-1}Q^{'}(\tau_{i-1};\gamma_{i-1})+\lambda_i Q^{'}(t;\gamma_i)\Bigg].\\
    E^{(\delta)}(\tau_l;\gamma_{l+1,l})&=\lambda_l Q(\tau_l;\gamma_l)-\lambda_{l+1} Q(\tau_l+\delta;\gamma_{l+1})\\
    &\qquad+\frac{\delta}{2}\Big\{\lambda_l Q^{'}(\tau_l;\gamma_l)+\lambda_{l+1} Q^{'}(\tau_l+\delta;\gamma_{l+1})\Big\}.
\end{align*}

Throughout the theoretical development and simulation study presented in this article, $\delta$ is assumed to be known.  However, in practical applications, $\delta$ is seldom available.  To address this issue, we briefly outline an approach based on the recommendations of Kannan and Kundu \cite{kannan2020simple}.  Their method offers a straightforward yet effective framework for estimating $\delta,$ ensuring the applicability of the theoretical results in real-world settings.

\subsubsection{Lag period $\delta$ unknown}\label{2.2.2}
In practical scenarios, $\delta$ may not always be known.  We estimate $\delta$, by maximizing profile likelihood function with respect to $\delta.$  As this maximization cannot be performed analytically, Kannan and Kundu \cite{kannan2020simple} suggested to employ grid search method.  For each candidate $\delta$ in the grid, MLE of the unknown parameters are computed as outlined in Subsection \ref{sec2.3} and corresponding likelihood value is evaluated.  The value of $\delta,$ yielding the maximum likelihood value is selected as the estimate for $\delta$.  This estimation procedure is applied exclusively in the data analysis of Section \ref{sec7}, where lag period $\delta$ is unknown in the lightbulb experiment conducted by Zhu \cite{zhu2010optimal}.

\subsection{\textbf{CRM under SSALT with interval monitoring}}\label{sec2.3}
\begin{figure}[htb!]%
\centering
\includegraphics[scale=0.32]{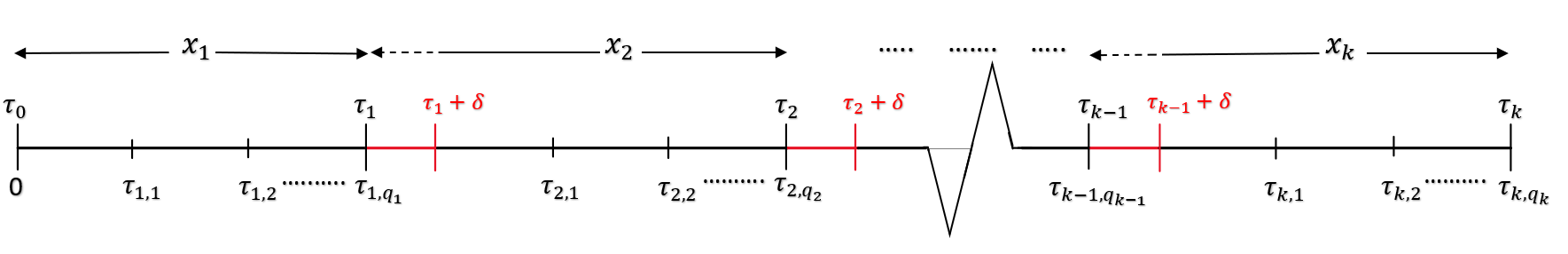}
\caption{SSALT under CRM with IMIIP.}\label{fig}
\end{figure}

As considered earlier, n NOSD is exposed to CRM SSALT experiment inspected at pre-fixed time points with the termination of the experiment at $\tau_k$.  Let $q_i$ be the number of inspection time points at stress level $x_i$ and $\tau_{i,m}$ be  $mth$ inspection time point at $ith$ stress level with $\tau_{i,q_i}=\tau_i\,;\,i=1,2,\dots,k\,;\,m=1,2,\dots,q_i\,;\,\tau_0=0$. 
 Figure \eqref{fig} depicts the layout of an SSALT CRM experiment with interval monitoring and intermediate inspection time points (IMIIP).
As the lag period consists of a very small amount of time, no failure is inspected during $(\tau_i,\tau_i+\delta)$.  Let us denote $n_{im}\,;\,i=1,2,\dots,k\,;\,m=1,2,\dots,q_i$ as number of observed failures in the interval $(\tau_{i(m-1)},\tau_{im}]$.  Then, $n_i=\sum_{i=1}^{q_i}n_{im}$ is total number of failures at $ith$ stress level and the total number of observed failures is thus given by $n_f=\sum_{i=1}^{k}n_i$.  Hence, $n_s=n-n_f$ is the number of survived units after time point $\tau_k.$  If $T$ is lifetime of a NOSD, then failure and survival probabilities using equation \eqref{sirv} are given as
\begin{align}
     p_{i1}&=P(\tau_{i-1}<T\leq \tau_{i,1})=\int_{\tau_{i-1}}^{\tau_{i-1}+\delta}f(x)\,dx+\int_{\tau_{i-1}+\delta}^{\tau_{i,1}}f(x)\,dx.\notag\\
     &=G^{(\delta)}(\tau_{i-1}+\delta;\gamma_{i-1,i})+\exp \left\{-\sum_{l=1}^{i-1}E^{(\delta)}(\tau_l;\gamma_{l+1,l})\right\} G^{(1)}(\tau_{i-1,i};\gamma_i).
    \end{align}
    \begin{align}
     p_{im}&=P(\tau_{i,m-1}<T\leq \tau_{i,m})\;;\;m=2,3,\dots,q_i.\notag\\
     &=\exp \left\{-\sum_{l=1}^{i-1}E^{(\delta)}(\tau_l;\gamma_{l+1,l})\right\} G^{(m)}(\tau_i;\gamma_i).\\
      p_s&=P(T>\tau_k)=\exp\left\{-\sum_{i=1}^{k-1}E^{(\delta)}(\tau_i;\gamma_{i+1,i})\right\} \exp\Big\{-\lambda_k Q(\tau_k;\gamma_k)\Big\},
\end{align}
where,
\begin{flalign*}
G^{(\delta)}(\tau_{i-1}+\delta;\gamma_{i-1,i})&=\exp\bigg[-\Big\{\lambda_{i-1}Q(\tau_{i-1};\gamma_{i-1})+\sum\limits_{l=1}^{i-2}E^{(\delta)}(\tau_l;\gamma_{l+1,l})\Big\}\bigg] \\
&\quad\qquad\bigg[1-\exp\Big\{-D^{(\delta)}(\tau_{i-1}+\delta;\gamma_{i-1,i})\Big\}\bigg].\\
 G^{(m)}(\tau_i;\gamma_i)&=\exp\Big\{-\lambda_i Q(\tau_{i,m-1};\gamma_i)\Big\}-\exp\Big\{-\lambda_i Q(\tau_{i,m};\gamma_i)\Big\} .\\
 G^{(1)}(\tau_{i-1,i};\gamma_i)&=\exp\Big\{-\lambda_i Q(\tau_{i-1}+\delta;\gamma_i)\Big\}-\exp\Big\{-\lambda_i Q(\tau_{i,1};\gamma_i)\Big\}.
 &&
\end{flalign*}
The log-likelihood function based on observed failure count data is given by 
\begin{equation*}
    ln\,L(\bm{\theta})\propto \left(\sum_{i=1}^{k}\sum_{m=1}^{q_i}n_{im}\,ln\,p_{im}\right)+\bigg(n_s\, ln\,p_s\bigg).
\end{equation*}
Therefore, maximum likelihood estimate (MLE) can be obtained as
\begin{equation*}
\hat{\bm{\theta}}=arg\mathop{max}_{\bm{\theta}}\,ln\, L(\bm{\theta})\;;\;\sum\limits_{i=1}^{k}\sum\limits_{m=1}^{q_i}n_{im}>0.
\end{equation*}

In the presence of outliers, MLE can give misinformation. Thus, a robust estimation method immune to small proportion of outliers is required.  The density power divergence (DPD) proposed by Basu et al.\cite{basu1998robust} is widely used for robust estimation \cite{baghel2024analysis,balakrishnan2024robust,baghel2024robust}.  Here, the DPD measure is computed as a divergence between empirical and theoretical failure (survival) probabilities for an NOSD testing unit.  For CRM under SSALT with IMIIP, the empirical failure and survival probabilities are given as $\left(\frac{n_{im}}{n},\frac{n_s}{n}\right)$ where $\,;\,i=1,2,\dots,k\,;\,m=1,2,\dots,q_i.$  Then, the DPD measure can be obtained as
\begin{align}
   D_{\alpha}(\bm{\theta})=\left\{p^{\alpha+1}_s+\sum_{i=1}^{k}\sum_{m=1}^{q_i}p^{\alpha+1}_{im}\right\}&-\left(1+\frac{1}{\alpha}\right)\left\{\frac{n_s}{n}p^{\alpha}_s+\sum_{i=1}^{k}\sum_{m=1}^{q_i}\frac{n_{im}}{n}p_{im}^{\alpha}\right\}\notag\\
   &\frac{1}{\alpha}\left\{\sum_{i=1}^{k}\sum_{m=1}^{q_i}\left(\frac{n_s}{n}\right)^{\alpha+1}+\sum_{i=1}^{k}\sum_{m=1}^{q_i}\left(\frac{n_{im}}{n}\right)^{\alpha+1}\right\},\label{dpd}
\end{align}
where $\alpha$ is termed as the tuning parameter. As $\alpha\to 0$, DPD measure approaches likelihood equation.  The minimum DPD estimator (MDE) can be obtained as
\begin{equation}
\hat{\bm{\theta}}_{\alpha}=arg\mathop{min}_{\bm{\theta}} D_{\alpha}(\bm{\theta}) .
\end{equation}
The asymptotic distribution of MDE is given in the appendix.  Further, with availability of prior knowledge, the Bayesian approach comes into the picture.  However, the conventional Bayesian approach relying on likelihood-based prior may yield unreliable estimates in the presence of outliers in data.  Thus, robust Bayesian estimation becomes necessary. 
 
\section{Robust Bayes method of estimation}\label{sec3}
Ghosh and Basu \cite{g2016} proposed to solve the non-robustness problem by replacing the likelihood function in the posterior with density power divergence (DPD) \cite{basu1998robust} based loss function, where the derived posterior is called a pseudo posterior.  We follow a similar approach for developing robust Bayesian inference for NOSD testing data. 
 The following subsection discusses the prior assumptions for the present study.

\subsection{\textbf{Prior assumptions}}
In Bayesian inference, the choice of prior governs the estimation.  As considered by Fan et al. \cite{fan2009bayesian}, we have taken prior information on $p_{im}$ instead of model parameters $\bm{\theta}$ for the first two prior selections.  To avoid a zero-frequency situation, we follow the idea of Lee and Morris \cite{le1985} and modify empirical probabilities as
\begin{equation} (\tilde{p}_{s}, \tilde{p}_{im})=\left(\frac{n_{s}+1}{n+k\sum_{i=1}^{k}q_i+1}, \frac{n_{im}+1}{n+k\sum_{i=1}^{k}q_i+1}\right), \label{emp}
\end{equation}
where, $i=1,2,\dots,k\;;\;m=1,2,\dots,q_i.$
\subsubsection{\textbf{Normal prior based on data}}
Assume $e_{im}$ is the error representing a difference between empirical and true failure probabilities.  Therefore, it can be expressed that 
\begin{equation}
\tilde{p}_{im}=p_{im}+e_{im}\;;\;i=1,2,\dots,k\;;\;m=1,2,\dots,q_i,\label{error}
\end{equation}
where, the error $e_{im}$ are assumed to be independent $N(0,\sigma^2)$ variables.  The conditional likelihood function as prior distribution of $\bm{\theta}$ given $\sigma^2$ can be obtained by  
\begin{equation*}
    L(\bm{\theta}\vert\sigma^2)\propto \prod_{i=1}^{k}\prod_{m=1}^{q_i}\frac{1}{\sigma\sqrt{2\pi}} \exp\left\{\frac{1}{2\sigma^2}(p_{im}-\tilde{p}_{im})^2\right\},
\end{equation*}
and $\pi(\sigma^2)\propto\frac{1}{\sigma^2}$ is the non-informative prior of $\sigma^2.$  The joint prior density of $\bm{\theta}$ can be obtained as
\begin{equation}
    \pi^{(Nor)}(\bm{\theta})\propto \int_{0}^{\infty} L(\bm{\theta}\vert\sigma^2)\pi(\sigma^2)\, d\bm{\theta}\propto \left\{\sum_{i=1}^{k}\sum_{m=1}^{q_i}(p_{im}-\tilde{p}_{im})^2\right\}^{-\sum_{i=1}^{k}q_i/2}.\label{norpri}
\end{equation}

\subsubsection{\textbf{Dirlichet prior based on data}}
Beta prior is a natural choice if a parameter can be interpreted as a probability.  Extending this idea, a Dirichlet prior is considered for the failure and survival probabilities as
\begin{equation}
    \pi^{(Dir)}(\bm{\theta})=\frac{p^{\beta_s-1}_s\prod_{i=1}^{k}\prod_{m=1}^{q_i}p^{\beta_{im}-1}_{im}}{Beta(\bm{\beta})},\label{bep}
\end{equation}
where, $\beta_s, \beta_{im}>0$ for $i=1,2,\dots,k\;;\;m=1,2,\dots,q_i$ and 
\begin{equation*}
Beta(\bm{\beta})=\frac{\Gamma\beta_s\prod_{i=1}^{k}\prod_{m=1}^{q_i}\Gamma\beta_{im}}{\Gamma(\beta_s+\sum_{i=1}^{k}\sum_{m=1}^{q_i}\beta_{im})}.
\end{equation*}
The hyper-parameters $\bm{\beta}$ are chosen such that
\begin{align}
    E(p_{im})&=\frac{\beta_{im}}{\beta_s+\sum_{i=1}^{k}\sum_{m^{'}=1}^{q_i}\beta_{im^{'}}}=\tilde{p}_{im}\;,\; E(p_s)= \frac{\beta_s}{\beta_s+\sum_{i=1}^{k}\sum_{m^{'}=1}^{q_i}\beta_{im^{'}}}=\tilde{p}_s\label{hyp1}\\
    Var(p_s)&=\frac{\beta_s\prod_{i=1}^{k}\prod_{m^{'}=1}^{q_i}\beta_{im^{'}}}{\left(\beta_s+\sum_{i=1}^{k}\sum_{m^{'}=1}^{q_i}\beta_{im^{'}}\right)^2\left(\beta_s+\sum_{i=1}^{k}\sum_{m^{'}=1}^{q_i}\beta_{im^{'}}+1\right)}=\sigma_{(p)}^2.\label{hyp2}
\end{align}
where, $\sigma_{(p)}^2$ is assumed to be a prefix quantity.  The estimates of hyper-parameters can be obtained by equations \eqref{hyp1} and \eqref{hyp2} as
\begin{align}
    \hat{\beta}_{im}&=\tilde{p}_{im}\left\{\frac{\tilde{p}_s(1-\tilde{p})}{\sigma_{(p)}^2}-1\right\}\;;\;
    \hat{\beta}_s=\left\{\frac{\tilde{p}_s(1-\tilde{p})}{\sigma_{(p)}^2}-1\right\}-\sum_{i=1}^{k}\sum_{m=1}^{q_i}\beta_{im}.\label{hypest}
\end{align}
Therefore, the joint prior density is given as
\begin{equation}
    \pi^{(Dir)}(\bm{\theta})\propto p^{\hat{\beta}_s-1}_s\prod_{i=1}^{k}\prod_{m=1}^{q_i}p^{\hat{\beta}_{im}-1}_{im}.\label{bepe}
\end{equation}
\subsubsection{\textbf{Order restricted prior assumption}}
The objective of an SSALT experiment is to accelerate the failure of the units by increasing the stress level.  Therefore, it is quite reasonable to assume that the expected lifetime of the experimental units is lower at the higher stress level.  Most of the inferences in the SSALT experiments ignore this assumption.  We incorporate this information by developing an order-restricted approach to the shape parameters, assuming an ordered Dirichlet-Gamma distribution as the joint prior.  For more information on order-restricted Bayes inference with this prior, refer to the studies \cite{mondal2020bayesian,pal2021bayesian,wiedner2021bayesian} and references therein.

If there is order restriction on shape parameters as $\gamma_i<\gamma_{i+1}\;;\;i=1,2,\dots,k-1$ and $\gamma=\sum_{i=1}^{k}\gamma_i$, then we assume that $$\gamma\sim GA(a_0,b_0)\;;\;\text{and}\; \bm{p}=\left(\frac{\gamma_i}{\gamma}\,,i=1,2,\dots,k\right)\sim DIR(\bm{a}),$$
where, $\bm{a}=(a_1,a_2,\dots,a_k).$  Then prior assumption on shape parameter takes the form of Dirichlet-Gamma pdf as\\
\begin{align}
\pi(\gamma_1,\gamma_2,\dots,\gamma_k|a_0,b_0,\bm{a})&=\frac{\Gamma{(a_1+a_2+\dots+a_k)}}{\Gamma a_0 \Gamma a_1\dots\Gamma a_k} b_0^{a_0}\gamma^{a_0-\sum_{i=1}^{k}a_i e^{-b_0\gamma}}\notag\\
&\qquad \sum_{\mathbb{P}}^{}\left(\gamma_{i_1}\gamma_{i_2}\dots\gamma_{i_k}\right)\;;\quad 0<\gamma_1< \gamma_2< \dots< \gamma_k<\infty.
\end{align}
Here, $\mathbb{P}$ denote the set of all $k!$ permutations on $\{1,2,\dots,k\}.$  The parameters $c_j$ assume Normal distribution as $c_j\sim N(\mu_j,\sigma_j^2)\;;\;j=0,1$.  Thus, joint prior distribution is given as
\begin{align}
\pi(c_0, c_1,\gamma_1,\gamma_2,\dots,\gamma_k|\bm{\theta}_H)&= \frac{1}{2\pi\sigma_0\sigma_1}\exp\left\{-\sum_{j=0}^{1}\frac{1}{2}\left(\frac{c_j-\mu_j}{\sigma_j}\right)^2\right\}\times\notag\\
&\quad\frac{\Gamma{(a_1+a_2+\dots+a_k)}}{\Gamma a_0 \Gamma a_1\dots\Gamma a_k} b_0^{a_0}\gamma^{a_0-\sum_{i=1}^{k}a_i} e^{-b_0\gamma}\times\notag\\
&\qquad \sum_{\mathbb{P}}^{}\left(\gamma_{i_1}^{a_1-1}\gamma_{i_2}^{a_2-1}\dots\gamma_{i_k}^{a_k-1}\right)\;;\; 0<\gamma_1< \gamma_2< \dots< \gamma_k<\infty, \label{orp}
\end{align}
where, $\bm{\theta}_H=\{\mu_j,\sigma_j, a_0, b_0,\bm{a}\}$ are the hyper parameters.

\subsection{\textbf{Posterior Analysis}}
For robust Bayesian inference in the context of NOSD, following the suggestion of Ghosh and Basu \cite{g2016}, a maximizer equation based on the DPD measure is presented as
\begin{equation}
    B_{\alpha}(\bm{\theta})=\frac{1}{\alpha}\left\{\frac{n_s}{n}p^{\alpha}_s+\sum_{i=1}^{k}\sum_{m=1}^{q_i}\frac{n_{im}}{n}p^{\alpha}_{im}\right\}-\frac{1}{\alpha+1}\left\{p_s^{\alpha+1}+\sum_{i=1}^{k}\sum_{m=1}^{q_i}p^{\alpha+1}_{im}\right\},\label{maxb}
\end{equation}
where, MDE with $\alpha>0$ is the maximizer of $B_{\alpha}(\bm{\theta})$.  Therefore, robust posterior density, a pseudo posterior, can be defined as
\begin{equation}
    \pi_{\alpha}(\bm{\theta}\vert data)=\frac{\exp(B_{\alpha}(\bm{\theta}))\pi(\bm{\theta})}{\int \exp(B_{\alpha}(\bm{\theta}))\pi(\bm{\theta}) d\bm{\theta}}.\label{psepost}
\end{equation}
Here, $\pi_{\alpha}(\bm{\theta}\vert data)$ is the proper density for $\alpha> 0.$  For $\alpha\to 0$, robust pseudo posterior will converge to conventional likelihood-based posterior density.  For NOSD testing data posterior densities are given as
\begin{itemize}
    \item Under Normal prior
   \begin{equation}
   \pi_{\alpha}^{(Nor)}(\bm{\theta}\vert data)\propto \exp(B_{\alpha}(\bm{\theta}))\left\{\sum_{i=1}^{k}\sum_{m=1}^{q_i}(p_{im}-\tilde{p}_{im})^2\right\}^{-\sum_{i=1}^{k}q_i/2}.\label{norpost}
\end{equation}
\item Under Dirichlet prior
\begin{equation}
   \pi_{\alpha}^{(Dir)}(\bm{\theta}\vert data)\propto \exp(B_{\alpha}(\bm{\theta}))\left\{p^{\hat{\beta}_s-1}_s\prod_{i=1}^{k}\prod_{m=1}^{q_i}p^{\hat{\beta}_{im}-1}_{im}\right\}.\label{dirpr}
\end{equation}
\item Under ordered restricted prior assumption
\begin{align}
 \pi_{\alpha}^{(Ord)}(\bm{\theta}\vert data)&\propto \exp\left\{B_{\alpha}(\bm{\theta})-\frac{1}{2}\Bigg(\left(\frac{c_0-\mu_0}{\sigma_0}\right)^2+\left(\frac{c_1-\mu_1}{\sigma_1}\right)^2\Bigg)\right\}\times \gamma^{a_0-\sum_{i=1}^{k}a_i} \times\notag\\
&\qquad e^{-b_0\gamma}\sum_{\mathbb{P}}^{}\left(\gamma_{i_1}\gamma_{i_2}\dots\gamma_{i_k}\right)\;;\; 0<\gamma_1< \gamma_2< \dots< \gamma_k<\infty.
\end{align} 
\end{itemize}
For any loss function $Loss(.,.)$, robust Bayes estimator (RBE) can be obtained as
\begin{equation*}
    arg  \min_t \int Loss(\bm{\theta}, t) \pi_{\alpha}(\bm{\theta}\vert data) d \bm{\theta}.
\end{equation*}
Particularly, for the squared error loss function, the robust Bayes estimator can be derived as
\begin{equation}
    \hat{\bm{\theta}}^{(b)}_{\alpha}=\int \bm{\theta} \pi_{\alpha}(\bm{\theta}\vert data) d \bm{\theta}.\label{rbe}
\end{equation}

Under all three prior assumptions, Bayes estimate cannot be obtained in closed form.  In such situations, Monte Carlo Markov Chain simulation methods can be used to approximate the Bayes estimates.  Since widely used methods like Gibbs sampler and Metropolis-Hastings (MH) algorithm struggle with high dimensional or highly correlated variables, therefore there has been a growing interest in using the Hamiltonian Monte Carlo (HMC) algorithm for Bayesian estimation recently \citep{thach2020improved,thomas2021learning,abba2024robust}.  The HMC steps are given in the algorithm \eqref{alg}.
\begin{algorithm}[htb!]
\caption{{\textbf{Hamiltonian Monte Carlo}}}\label{alg}
\begin{itemize}
\item Define the diagonal matrix $\bm{M}$, step size $\epsilon$, leapfrog step $L$ and sample size $N$.
\item Initialize the position state $\bm{\Lambda}^{(0)}.$
\item[] For $t=1,2,\dots,N$
\item Sample $\bm{\phi}^{(t)}\sim N(\bm{0},\bm{M}).$
\item Run leapfrog starting at $(\bm{\Lambda}^{(t)},\bm{\phi}^{(t)})$ for $L$ step with step size $\epsilon$ to produce proposed state $(\bm{\Lambda}^{*},\bm{\phi}^{*})$.
\item[] Let $\bm{\phi}^{(t,0)}=\bm{\phi}^{(t)}$ and  $\bm{\Lambda}^{(t-1,0)}=\bm{\Lambda}^{(t-1)}$, then for $t^{'}=1,2,\dots,N$
\item $\bm{\phi}_{\epsilon/2}=\bm{\phi}^{(t,t^{'}-1)}+\left.\frac{\epsilon}{2}\frac{\partial \log\pi_{\alpha}(\bm{\Lambda}\vert t)}{\partial\bm{\Lambda}}\right\vert_{\bm{\Lambda}=\bm{\Lambda}^{(t-1,t^{'}-1)}}$
\item $\bm{\Lambda}^{t-1,t^{'}}=\bm{\Lambda}^{t-1,t^{'}-1}+\epsilon\, \bm{M}^{-1}\bm{\phi}_{\epsilon/2}$
\item $\bm{\phi}^{t,t^{'}}=\bm{\phi}_{\epsilon/2}+\left.\frac{\epsilon}{2}\frac{\partial \log\pi_{\alpha}(\bm{\Lambda}\vert t)}{\partial\bm{\Lambda}}\right\vert_{\bm{\Lambda}=\bm{\Lambda}^{(t-1,t^{'})}}$
\item[] Hence, $\bm{\Lambda}=\bm{\Lambda}^{t-1,L}$ and $\phi^{*}=\phi^{t,L}$.
\item Compute acceptance probability\\ $acc=min\Big\{1,\exp\big(U(\bm{\Lambda}^{(t-1)})-U(\bm{\Lambda}^{*})+K(\bm{\phi}^{(t)})\big)-K(\bm{\phi}^{*})\big)\Big\},$\\
where, $U(\bm{\Lambda})=-log\,\pi_{\alpha}(\bm{\theta})$ and $K(\phi)=\frac{1}{2}\phi^T\bm{M}^{-1}\phi.$
\item Generate a random number $u\sim U(0,1)$ and set\\
$\bm{\theta}^{(t)}=\begin{cases}
\bm{\theta}^{*}&; \;u\leq acc.\\
\bm{\theta}^{(t-1)}&;\; \text{otherwise}.
\end{cases} $
\item Stop when $t=N$.
\end{itemize}
\end{algorithm}
In the HMC algorithm, $m^{'}$ chains of $N$ values each are generated, and the first $N_{0}$ values are discarded as a burn-in period in each chain.  A total $N^{'}=N-N_0$ values for each parameter are finally obtained.  Based on these obtained values, the Bayes estimates and the highest posterior density credible intervals (HPD CRI) of the model parameters can be approximated using algorithm \eqref{alcred}.
\begin{algorithm}[htb!]
	\caption{{Bayes Estimates and HPD Credible Intervals}}\label{alcred}
\begin{itemize}
\item The Bayes estimator based on squared error loss can be approximated  as
$\bm{\hat{\theta}}=\frac{1}{m^{'}N^{'}}\sum_{l=1}^{m^{'}}\sum_{t=N_{0}+1}^{N}\bm{\theta}^{(t)}_{l},$
where $\bm{\theta}^{(t)}_{l}$ is the value of $t^{th}$ iteration for $l^{th}$ chain.
		\item For $100(1-\xi)\%$ CRI of $\bm{\theta}$:\\
		Sort $\bm{\theta}^{(i)}$'s in ascending order to obtain $(\bm{\theta}^{(1)}, \bm{\theta}^{(2)}, \ldots, \bm{\theta}^{(N^{'})})$ and 
		$(\bm{\theta}^{(j)},\bm{\theta}^{(j+[N^{'}(1-\xi)])})$ for $ j=1,\dots,[N^{'}\xi]$ is the $100(1-\xi)\%$ credible intervals.
		\item The $100(1-\xi)\%$ HPD CRI is $(\bm{\theta}^{(j^*)},\bm{\theta}^{(j^*+[N^{'}(1-\xi)])})$ such that 
		$(\bm{\theta}^{(j^*)},\bm{\theta}^{(j^*+[N^{'}(1-\xi)])})\leq (\bm{\theta}^{(j)},\bm{\theta}^{(j+[N^{'}(1-\xi)])}); j=1,\dots,[N^{'}\xi]$.
	\end{itemize}
\end{algorithm}

\section{Testing of hypothesis based on robust Bayes factor}\label{sec4}
Validating whether available data supports the hypothesis of interest is essential for inferential study.  For datasets with outliers, robust testing of the hypothesis is pertinent.  This section develops robust testing of the hypothesis based on the Bayes factor inspired by the procedure followed by Ghosh et al. \cite{ghosh2006introduction}.  For parameter $\bm{\theta}=(c_0, c_1, \gamma_i\;;\;i=1,2,\dots,k),$ consider the vector-valued function $fn:\mathbb{R}^{k+2}\xrightarrow[]{}\mathbb{R}^{w}.$  The null and alternative hypotheses are given as  
\begin{equation*}
    \bm{H}_0 : \bm{\theta}\in\bm{\Theta}_0\quad \text{against}\quad \bm{H}_1 : \bm{\theta}\in\bm{\Theta}_1,
\end{equation*}
where, $\bm{\Theta}_0=\{\bm{\theta}\in\bm{\Theta}_0 : fn(\bm{\theta})=\bm{0}_w\}$ and $\bm{\Theta}_1=\{\bm{\theta}\notin\bm{\Theta}_0\}$.  Let $\rho_0$ and $1-\rho_0$ be prior probabilities under $\bm{\Theta}_0$ and $\bm{\Theta}_1$ respectively.  Let $\pi_j(\bm{\theta})$ be prior density of $\bm{\theta}$ under $\bm{\Theta}_j$ such that, $\int_{\bm{\Theta}_j}\pi_j(\bm{\theta})d\bm{\theta}=1\,;\,j=0,1$.  Then, the prior can be expressed as
\begin{equation*}
    \pi(\bm{\theta})=\rho_0\pi_0(\bm{\theta}) I\{\bm{\theta}\in\bm{\Theta}_0\}+(1-\rho_0)\pi_1(\bm{\theta}) I\{\bm{\theta}\in\bm{\Theta}_1\}.
\end{equation*}
Therefore, posterior probabilities under $\bm{\Theta}_0$ and $\bm{\Theta}_1$ are
\begin{align*}
    P_{\pi_{\alpha}}(\bm{\theta}\in\bm{\Theta}_0\vert data)&=\frac{\rho_0}{M_{\alpha}(\pi)}\int_{\bm{\Theta}_0}\exp(B_{\alpha}(\bm{\theta}))\pi_0(\bm{\theta})\,d\bm{\theta}.\\
     P_{\pi_{\alpha}}(\bm{\theta}\in\bm{\Theta}_1\vert data)&=\frac{(1-\rho_0)}{M_{\alpha}(\pi)}\int_{\bm{\Theta}_1}\exp(B_{\alpha}(\bm{\theta}))\pi_1(\bm{\theta})\,d\bm{\theta},
\end{align*}
where, $M_{\alpha}(\pi)$ is the marginal density expressed as 
\begin{equation*}
M_{\alpha}(\pi)=\rho_0\int_{\bm{\Theta}_0}\exp(B_{\alpha}(\bm{\theta}))\pi_0(\bm{\theta})d\bm{\theta}+(1-\rho_0)\int_{\bm{\Theta}_1}\exp(B_{\alpha}(\bm{\theta}))\pi_1(\bm{\theta})\,d\bm{\theta}.
\end{equation*}
The posterior odds ratio of $H_0$ relative to $H_1$ is given as
\begin{equation}
   \frac{ P_{\pi_{\alpha}}(\bm{\theta}\in\bm{\Theta}_0\vert data)}{ P_{\pi_{\alpha}}(\bm{\theta}\in\bm{\Theta}_1\vert data)}=\left(\frac{\rho_0}{1-\rho_0}\right)BF_{01},\label{odds}
\end{equation}
where, $BF_{01}$ is the Bayes factor given as
\begin{equation}
    BF_{01}=\frac{\int_{\bm{\Theta}_0}\exp(B_{\alpha}(\bm{\theta}))\,d\bm{\theta}}{\int_{\bm{\Theta}_1}\exp(B_{\alpha}(\bm{\theta}))\,d\bm{\theta}}.\label{bysf}
\end{equation}
The Bayes factor measures the strength of evidence the data offers supporting one hypothesis over another. Jeffreys \cite{jeffreys1998theory} suggested a scale to interpret the Bayes factor, and Kass and Raftery \cite{kass1995bayes} simplified it further, which is given in Table \ref{tab1}.
\begin{table}[htb!]
\caption{Interpretation of Bayes factor \citep{kass1995bayes}.}\label{tab1}%
\begin{tabular}{@{}ll@{}}
\toprule
\textbf{BF$_{01}$}&\textbf{Support for $\bm{H}_0$}  \\
\midrule
$<1$&Negative\\
1 to 3& Not worth more than a bare mention\\
3 to 20 & Positive\\
20 to 150 & Strong\\
$>150$ & Very Strong\\
\botrule
\end{tabular}
\end{table}

\section{Property of robustenss}\label{sec5}
This section includes robustness analysis through influence function (IF).  Suppose, for a true distribution $F_{\bm{\theta}}$, functional of any estimator is denoted by $T_{\alpha}(F_{\bm{\theta}}).$  Then, the influence function is defined as
\begin{equation}
    IF(t;T_{\alpha},F_{\bm{\theta}})=\lim_{\epsilon \to 0}\frac{T_{\alpha}(U_{\varepsilon})-T_{\alpha}(F_{\bm{\theta}})}{\epsilon}=\left.\frac{\partial(T_{\alpha}(U_{\epsilon}))}{\partial\epsilon}\right\vert_{\epsilon\to 0^{+}}.
\end{equation}
Here, $U_{\epsilon}=(1-\epsilon)F_{\bm{\theta}}+\epsilon\Delta_t$ is the contaminated model where $\epsilon,$ $(0<\epsilon<1)$ is the proportion of contamination and $\Delta_t$ denotes the degenerate distribution at point $t$.  

Let $F_{\bm{\theta}}$ be the true distribution from where data is generated.  If $T_{\alpha}(F_{\bm{\theta}})$ is statistical functional of MDE $\hat{\bm{\theta}}_{\alpha}$, $T_{\alpha}(F_{\bm{\theta}})$ will be the value of $\bm{\theta}$ which will minimize,
\begin{equation}
  p^{\alpha+1}_s+\sum_{i=1}^{k}\sum_{m=1}^{q_i}p^{\alpha+1}_{im}-\left(1+\frac{1}{\alpha}\right)\left\{\left(\int_{I_s}dF_{\bm{\theta}}\right)p^{\alpha}_s+\sum_{i=1}^{k}\sum_{m=1}^{q_i}\left(\int_{I_{im}}dF_{\bm{\theta}}\right)p^{\alpha}_{im}\right\},\label{mizdp}
\end{equation}
where $I_s=\{t: t>\tau_k \}$ and $I_{im}=\{ t: \tau_{i,m-1}<t\leq\tau_{i,m} \}$.  Replacing $F_{\bm{\theta}}$ by contaminated model $U_{\epsilon}=(1-\epsilon)F_{\bm{\theta}}+\epsilon\Delta_t$ in equation \eqref{mizdp} and differentiating it with respect to $\epsilon$ and putting $\epsilon\to 0^{+}$, we get The influence function of $\hat{\bm{\theta}}_{\alpha}$ for NOSD testing units as 
 \begin{flalign}
    IF(t;T_{\alpha},F_{\bm{\theta}})=J^{-1}_{\alpha}(\bm{\theta})&\left[\left\{\Delta^{(I_s)}_t-p_s\right\}p^{\alpha-1}_s\frac{\partial(p_s)}{\partial\bm{\theta}}\right.\notag\\
    &\qquad\left.+\sum_{i=1}^{k}\sum_{m=1}^{q_i}\left\{\Delta^{(I_{im})}_t-p_{im}\right\}p^{\alpha-1}_{im}\frac{\partial(p_{im})}{\partial\bm{\theta}}\right].
    &&
\end{flalign}
Here,
$\Delta^{(I)}_t = 
\begin{cases}
1 \quad \text{if} \  t \in I\\
0 \quad \text{otherwise}.\\
\end{cases}
$ \\

\subsection{\textbf{Influence function of RBE}}
To study robustness through IF \citep{g2016}, Bayes functional of $\hat{\bm{\theta}}^{(b)}_{\alpha}$ under squared error loss function is given as
\begin{equation}
    T^{(b)}_{\alpha}(F_{\bm{\theta}})=\frac{\int\bm{\theta}\exp\big\{B_{\alpha}(\bm{\theta};F_{\bm{\theta}})\big\}\pi(\bm{\theta})\,d\bm{\theta}}{\int \exp\big\{B_{\alpha}(\bm{\theta};F_{\bm{\theta}})\big\}\pi(\bm{\theta})\,d\bm{\theta}},\label{funct}
\end{equation}
\begin{flalign*}
\text{where, }B_{\alpha}(\bm{\theta};F_{\bm{\theta}})&=\frac{1}{\alpha}\left\{\left(\int_{I_s}dF_{\bm{\theta}}\right)p^{\alpha}_s(\bm{\theta})+\sum_{i=1}^{k}\sum_{m=1}^{q_i}\left(\int_{I_{im}}dF_{\bm{\theta}}\right)p^{\alpha}_{im}(\bm{\theta})\right\}\\
&\qquad\qquad-\frac{1}{\alpha+1}\left\{p^{\alpha+1}_s(\bm{\theta})+\sum_{i=1}^{k}\sum_{m=1}^{q_i}p^{\alpha+1}_{im}(\bm{\theta})\right\}.
    &&
\end{flalign*}
\begin{res}\label{res5}
The influence function of Bayes estimator $\hat{\bm{\theta}}^{(b)}_{\alpha}$ under for NOSD testing units is given by
\begin{equation*}
IF(t;T^{(b)}_{\alpha},F_{\bm{\theta}})=Cov_{(p)}(\bm{\theta},X_{\alpha}(\bm{\theta};t,f_{\bm{\theta}})),
	\end{equation*}
where, $Cov_{(p)}()$ is the covariance for posterior distribution and
\begin{flalign*}
    X_{\alpha}
    &=\frac{1}{\alpha}\left[\left\{\Delta^{(I_s)}_t-p_s(\bm{\theta})\right\}p^{\alpha}_s(\bm{\theta})+\sum_{i=1}^{k}\sum_{m=1}^{q_i}\left\{\Delta^{(I_{im})}_t-p_{im}(\bm{\theta})\right\}p^{\alpha}_{im}(\bm{\theta})\right].
    &&
\end{flalign*}
\end{res}
\begin{proof}
Given in the appendix.
\end{proof}

\subsection{\textbf{Influence function of Bayes factor}}
Here, the robustness property of the Bayes factor is examined by deriving its IF when the null hypothesis is true.  Let $F_{\theta_0}$ be the true distribution under the null hypothesis $H_0: \bm{\theta}\in\bm{\Theta}_0$  and therefore functional related to the Bayes factor can be defined as
\begin{equation}
    T^{(\alpha)}_{\bm{\Theta}}(F_{\bm{\theta}_0})=\frac{\int_{\bm{\Theta}_0}\exp\big\{B_{\alpha}(\bm{\theta}\in\bm{\Theta}_0;F_{\bm{\theta}_0})\big\}\pi_0(\bm{\theta})\,d\bm{\theta}}{\int_{\bm{\Theta}_1}\exp\big\{B_{\alpha}(\bm{\theta}\in\bm{\Theta}_1;F_{\bm{\theta}_1})\big\}\pi_1(\bm{\theta})\,d\bm{\theta}}.
\end{equation}
Here, $B_{\alpha}(\bm{\theta}\in\bm{\Theta}_j;F_{\bm{\theta}_0})\,;\,j=0,1$, is expressed as
\begin{flalign*}
B_{\alpha}(\bm{\theta}\in\bm{\Theta}_j;F_{\bm{\theta}_0})&=\frac{1}{\alpha}\left\{\left(\int_{I_s}dF_{\bm{\theta}_0}\right)p^{\alpha}_s(\bm{\theta}\in\bm{\Theta}_j)+\sum_{i=1}^{k}\sum_{m=1}^{q_i}\left(\int_{I_{im}}dF_{\bm{\theta}_0}\right)p^{\alpha}_{im}(\bm{\theta}\in\bm{\Theta}_j)\right\}\\
&\qquad\qquad-\frac{1}{\alpha+1}\left\{p^{\alpha+1}_s(\bm{\theta}\in\bm{\Theta}_j)+\sum_{i=1}^{k}\sum_{m=1}^{q_i}p^{\alpha+1}_{im}(\bm{\theta}\in\bm{\Theta}_j)\right\}.
    &&
\end{flalign*}
Let contamination in the true distribution $F_{\theta_0}$ under $H_0: \bm{\theta}\in\bm{\Theta}_0$ be given as $U_{\epsilon}=(1-\epsilon)F_{\bm{\theta}_0}+\epsilon\Delta_t$. Then, the following result provides an explicit expression of IF under the given setup.
\begin{res}\label{res6}
The influence function of Bayes factor BF$_{01}$ is obtained as
\begin{equation*}
IF(t;T^{(\alpha)}_{\bm{\Theta}},F_{\bm{\theta}_0})=Y_{\alpha}(\bm{\Theta})\bigg\{E\Big[X_{\alpha}(\bm{\theta}\in\bm{\Theta}_0)\Big]-E\Big[X_{\alpha}(\bm{\theta}\in\bm{\Theta}_1)\Big]\bigg\},
\end{equation*}
where,
\begin{equation*}
    Y_{\alpha}(\bm{\Theta})=\frac{\int_{\bm{\Theta}_0}\exp\big\{B_{\alpha}(\bm{\theta}\in\bm{\Theta}_0)\big\}\pi_0(\bm{\theta})\,d\bm{\theta}}{\int_{\bm{\Theta}_1}\exp\big\{B_{\alpha}(\bm{\theta}\in\bm{\Theta}_1)\big\}\pi_1(\bm{\theta})\,d\bm{\theta}}.
\end{equation*}
\end{res}
\begin{proof}
Given in the appendix.
\end{proof}
The maximum value of IF shows the degree of bias resulting from contamination.  Therefore, the smaller the value of IF, the more robust the estimator or Bayes factor.  Also, for all the estimators and Bayes factor, IF is a bounded function of $t.$

\section{Simulation study}\label{sec6}
For simulation analysis, $50$ NOSD testing units are put into three-step-stress ALT with $m=9$ inspection time points under a cumulative risk model with a known lag period $\delta=0.1.$  Stress levels and stress change time points are taken as $x_1{=}5,\tau_1{=}3;\,x_2{=}6,\tau_2{=}12;\,x_3{=}7,\tau_3{=}21$ with some suitable units.  The experiment is terminated at $\tau_3=21.$ The intermediate inspection times are set as $(0.1,1.5,3,6,9,12,15,18,21).$  The failed NOSD units are counted at each of the inspection times and number of survived units are recorded after termination time point $\tau_3=21.$  
\begin{table}[htb!]
\caption{Information regarding simulation.}\label{intab}%
\begin{tabular}{@{}lll@{}}
\toprule
\textbf{Information}&\textbf{Weibull}&\textbf{Gompertz}\\
\midrule
True parameters&$(c_0{=}{-}0.52,c_1{=}{-}0.1,$&$(c_0{=}0.01,c_1{=}-0.1,$\\
& $\gamma_1{=}0.1,\gamma_2{=}0.4,\gamma_3{=}0.8)$&$\gamma_1{=}0.05,\gamma_2{=}0.06,\gamma_3{=}0.08$\\
Outlying parameters& $(c_0{-}0.13,c_1{-}0.05,$ &$(c_0{+}0.005,c_1{+}0.02,$\\
&$\gamma_1{+}0.11,\gamma_2{+}0.12,\gamma_3{+}0.13)$&$\gamma_1{-}0.005,\gamma_2{+}0.007,\gamma_3{+}0.01)$\\
Dirichlet prior&$\sigma_{(p)}^2=0.05$&$\sigma_{(p)}^2=0.05$\\
\multicolumn{3}{@{}l@{}}{Hamiltonian Monte Carlo}\\
Step (size, no.)&$(\epsilon,L){=}(0.02,8)$&$(\epsilon,L){=}(0.01,7)$\\
$M{=}1/v$&$v{=}(1,1,1,1,1)$&$v{=}(1,1,1,1,1)$ \\
\botrule
\end{tabular}
\end{table}
To generate data from lifetime distribution under the given set-up, true model parameters are set as $\bm{\theta}{=}(c_0,c_1,\gamma_1,\gamma_2,\gamma_3)^{'}.$  To study robustness, we incorporate contamination by deviating the failure mechanism of NOSD testing units from the assumed model.  The idea is similar to the contamination scheme adopted in the studies of \cite{balakrishnan2019robust,balakrishnan2019robust1,balakrishnan2019robust2,balakrishnan2020robust,baghel2024analysis,baghel2024robust}.  The contamination is employed by deviating the parameters from true parameters, which results in deviation of failure mechanism from assumed model.  The information regarding the simulation study for two special cases of the Lehman family, namely Weibull and Gompertz lifetime distribution, is provided in Table \eqref{intab}.  

\begin{table}[htb!]
\caption{Bias of the estimates with Weibull lifetime distribution.}\label{tab2a}%
\setlength{\tabcolsep}{2.5pt} 
\begin{tabular}{@{}lllllllllll@{}}
\toprule
& \multicolumn{5}{@{}c@{}}{\textbf{Pure data}}\\
\cmidrule(lr){2-6}
&$\bm{\hat{c}}_0$&$\bm{\hat{c}}_1$&$\bm{\hat{\gamma}}_1$&$\bm{\hat{\gamma}}_2$&$\bm{\hat{\gamma}}_3$\\
\midrule
MLE&  -0.008209& -0.009051&  0.008961&  0.009917 & 0.013870 \\
MDE$_{\alpha{=}0.3}$&-0.020796& -0.009768&  0.013747&  0.010794&  0.021681   \\
MDE$_{\alpha{=}0.9}$&  0.022718& -0.011962& -0.009301& -0.010995 &0.010578\\
BE$^{(Nor)}$& -0.004994 &{-0.004820}&\textbf{-0.004456} &\textbf{-0.004467} & 0.005059 \\
BE$^{(Dir)}$&{-0.004185}& \textbf{-0.004834}&-0.005135 & -0.004715 & \textbf{0.003951} \\
BE$^{(Ord)}$& -0.004978& -0.004944&-0.004889& -0.004827 & 0.005015 \\
RBE$^{(Nor)}_{\alpha{=}0.3}$& \textbf{-0.003317}& -0.005772& -0.004458& -0.005002 & 0.005825 \\
RBE$^{(Nor)}_{\alpha{=}0.9}$&-0.005258 & -0.004872 & -0.004596 &-0.004816   &0.005143 \\
RBE$^{(Dir)}_{\alpha{=}0.3}$& -0.004108 &-0.006574& -0.005213& -0.005681 & 0.005458 \\
RBE$^{(Dir)}_{\alpha{=}0.9}$& -0.005197  &-0.005607&-0.005545& -0.005304 & 0.005806\\
RBE$^{(Ord)}_{\alpha{=}0.3}$&-0.004857& -0.005064& -0.004909& -0.005105 & 0.005126  \\
RBE$^{(Ord)}_{\alpha{=}0.9}$&-0.005117  &-0.005063& -0.004848& -0.005101&  0.004826\\
\midrule
& \multicolumn{5}{@{}c@{}}{\textbf{Contamination}}\\\cmidrule(lr){2-6}
&$\bm{\tilde{c}}_0$&$\bm{\tilde{c}}_1$&$\bm{\tilde{\gamma}}_1$&$\bm{\tilde{\gamma}}_2$&$\bm{\tilde{\gamma}}_3$\\
\midrule
MLE&-0.064410& -0.026852&  0.049347&  0.092497&  0.036472\\
MDE$_{\alpha{=}0.3}$&-0.054801& -0.030212 & 0.034207 & 0.052207&  0.027917 \\
MDE$_{\alpha{=}0.9}$&-0.039971& -0.019991 & 0.020333  &0.006092 & 0.020596\\
BE$^{(Nor)}$& -0.010427& -0.010053 & 0.014594 &\textbf{-0.003682} & 0.009002 \\
BE$^{(Dir)}$& -0.009591& -0.009590&  0.012659& -0.003723 & 0.010586 \\
BE$^{(Ord)}$&-0.010109& -0.008003 & 0.010600& -0.003864&  0.010063 \\
RBE$^{(Nor)}_{\alpha{=}0.3}$ &-0.005966 &-0.006317& -0.005811& -0.005530 & 0.006035\\
RBE$^{(Nor)}_{\alpha{=}0.9}$ & -0.005884& -0.005243& -0.005320& -0.005152 & 0.005098 \\
RBE$^{(Dir)}_{\alpha{=}0.3}$&\textbf{-0.004686}& -0.006761& -0.005996& -0.006223 & 0.005966\\
RBE$^{(Dir)}_{\alpha{=}0.9}$& -0.005996 &-0.005692& -0.004917& -0.005639 & 0.005968\\
RBE$^{(Ord)}_{\alpha{=}0.3}$&-0.005032& \textbf{-0.005086}& \textbf{-0.004948}& -0.004921 & 0.005218\\
RBE$^{(Ord)}_{\alpha{=}0.9}$&-0.005392& -0.005289& -0.004989& -0.005712&  \textbf{0.005060} \\
\botrule
\end{tabular}
\end{table}

\begin{table}[htb!]
\caption{RMSE of the estimates with Weibull lifetime distribution.}\label{tab2b}%
\setlength{\tabcolsep}{2.5pt} 
\begin{tabular}{@{}lllllllllll@{}}
\toprule
& \multicolumn{5}{@{}c@{}}{\textbf{Pure data}}\\\cmidrule(lr){2-6}
&$\bm{\hat{c}}_0$&$\bm{\hat{c}}_1$&$\bm{\hat{\gamma}}_1$&$\bm{\hat{\gamma}}_2$&$\bm{\hat{\gamma}}_3$\\
\midrule
MLE
& 0.077198& 0.025678& 0.058675& 0.090243& 0.046611 \\
MDE$_{\alpha{=}0.3}$&0.210376& 0.014949& 0.113615& 0.086481 &0.058808
\\
MDE$_{\alpha{=}0.9}$
& 0.089629& 0.011963& 0.009482& 0.010995& 0.011744\\
BE$^{(Nor)}$
& 0.017092 & 0.013913 & 0.010930 & 0.011037 & 0.014518 \\
BE$^{(Dir)}$
& 0.017085 & 0.014667 & 0.011517 & 0.011395 & 0.013630 \\
BE$^{(Ord)}$
& \textbf{0.005424} & \textbf{0.005551} & \textbf{0.005486} & \textbf{0.005355} & 0.005640 \\
RBE$^{(Nor)}_{\alpha{=}0.3}$&0.020105 &0.017666& 0.014384 &0.014292& 0.017668
 \\
RBE$^{(Nor)}_{\alpha{=}0.9}$
& 0.017119 & 0.014342 & 0.011182 & 0.011496 & 0.014752  \\
RBE$^{(Dir)}_{\alpha{=}0.3}$&0.019869& 0.017817& 0.014505 &0.013990 &0.017164
 \\
RBE$^{(Dir)}_{\alpha{=}0.9}$
& 0.017841 & 0.014758 & 0.011767 & 0.011006 & 0.015121 \\
RBE$^{(Ord)}_{\alpha{=}0.3}$&0.005347& 0.005612& 0.005526& 0.005704& 0.005759
& \\
RBE$^{(Ord)}_{\alpha{=}0.9}$
&  0.005585 & 0.005606 & 0.005453 & 0.005622 & \textbf{0.005406} \\
\midrule
& \multicolumn{5}{@{}c@{}}{\textbf{Contamination}}\\\cmidrule(lr){2-6}
&$\bm{\tilde{c}}_0$&$\bm{\tilde{c}}_1$&$\bm{\tilde{\gamma}}_1$&$\bm{\tilde{\gamma}}_2$&$\bm{\tilde{\gamma}}_3$\\
\midrule
MLE
& 0.248244& 0.039385& 0.186421& 0.145880& 0.079131 \\
MDE$_{\alpha{=}0.3}$
& 0.121825& 0.033286 &0.130116 &0.094116& 0.065120\\
MDE$_{\alpha{=}0.9}$
& 0.039971& 0.019991 &0.020336& 0.011306& 0.020602 \\
BE$^{(Nor)}$
&0.023125& 0.013018& 0.020611& 0.010673& 0.021102 \\
BE$^{(Dir)}$
&0.022439& 0.012913& 0.020513& 0.010760& 0.020689 \\
BE$^{(Ord)}$
& 0.010147& 0.009012& 0.007517 &0.006119 &0.008206 \\
RBE$^{(Nor)}_{\alpha{=}0.3}$
& 0.021102& 0.017382 &0.015226& 0.014241 &0.017544 \\
RBE$^{(Nor)}_{\alpha{=}0.9}$
& 0.021010 &0.017744& 0.014549& 0.014557& 0.016929 \\
RBE$^{(Dir)}_{\alpha{=}0.3}$
& 0.020887& 0.017774& 0.014705& 0.013953& 0.017742 \\
RBE$^{(Dir)}_{\alpha{=}0.9}$
& 0.020387& 0.017663 &0.014304& 0.014242& 0.017295 \\
RBE$^{(Ord)}_{\alpha{=}0.3}$
& \textbf{0.005402}& 0.005540& \textbf{0.005352}& \textbf{0.005352}& 0.005519 \\
RBE$^{(Ord)}_{\alpha{=}0.9}$
&0.005435& \textbf{0.005486}& 0.005467& 0.005506& \textbf{0.005515}  \\
\botrule
\end{tabular}
\end{table}

\begin{table}[htb!]
\caption{Bias of the estimates with Gompertz lifetime distribution.}\label{tab3a}%
\setlength{\tabcolsep}{2.5pt} 
\begin{tabular}{@{}lllllllllll@{}}
\toprule
& \multicolumn{5}{@{}c@{}}{\textbf{Pure data}}\\\cmidrule(lr){2-6}
&$\bm{\hat{c}}_0$&$\bm{\hat{c}}_1$&$\bm{\hat{\gamma}}_1$&$\bm{\hat{\gamma}}_2$&$\bm{\hat{\gamma}}_3$\\
\midrule
$\textbf{MLE}$ & 0.004309 & 0.018255& -0.002098&  0.003723 & 0.001384\\ 
$\textbf{MDE}_{\alpha{=}0.3}$&  0.007239&  0.014833& -0.002121&  0.003159&  0.001722  \\
$\textbf{MDE}_{\alpha{=}0.9}$& 0.008997 & 0.015477 &-0.004287&-0.007593&  0.000274\\
$\textbf{BE}^{(Nor)}$& 0.000222& -0.001061&  0.000055&  0.000134& 0.000290 \\
$\textbf{BE}^{(Dir)}$& -0.000007& -0.001035&  0.000163&  0.000129 &0.000352\\
$\textbf{BE}^{(Ord)}$& 0.000044& \textbf{-0.000979}&  0.000132& \textbf{0.000046} & 0.000265\\
$\textbf{RBE}^{(Nor)}_{\alpha{=}0.3}$&-0.000352& -0.001401& -0.000273& -0.000642 & 0.000402  \\
$\textbf{RBE}^{(Nor)}_{\alpha{=}0.9}$&  0.000091& -0.001034 & 0.000157 & 0.000127 & 0.000316 \\
$\textbf{RBE}^{(Dir)}_{\alpha{=}0.3}$&-0.000206& -0.001039&  0.000708& -0.000216 & 0.000769  \\
$\textbf{RBE}^{(Dir)}_{\alpha{=}0.9}$& 0.000168& -0.001046&  \textbf{0.000044}&  0.000179 & 0.000307 \\
$\textbf{RBE}^{(Ord)}_{\alpha{=}0.3}$&-0.000063& -0.001068&  0.000463& -0.000155&  0.000519 \\
$\textbf{RBE}^{(Ord)}_{\alpha{=}0.9}$& \textbf{-0.000006}& -0.001018 & 0.000171&  0.000198 &\textbf{0.000226}\\
\midrule
& \multicolumn{5}{@{}c@{}}{\textbf{Contamination}}\\\cmidrule(lr){2-6}
&$\bm{\tilde{c}}_0$&$\bm{\tilde{c}}_1$&$\bm{\tilde{\gamma}}_1$&$\bm{\tilde{\gamma}}_2$&$\bm{\tilde{\gamma}}_3$\\
\midrule
$\textbf{MLE}$&0.013019&  0.038365& -0.017612 & 0.014066&  0.003366\\
$\textbf{MDE}_{\alpha{=}0.3}$&0.009514 & 0.019116 &-0.004752&  0.009021 & 0.000505\\
$\textbf{MDE}_{\alpha{=}0.9}$&0.009128&  0.016598& -0.004802& -0.008086&  0.000271 \\
$\textbf{BE}^{(Nor)}$&0.001093& -0.010038&  0.000968&  0.001043&  0.000997 \\
$\textbf{BE}^{(Dir)}$&0.001025& -0.009979 & 0.001057 & 0.000875&  0.001019\\
$\textbf{BE}^{(Ord)}$&0.000967& -0.010051 & 0.001034&  0.001024 & 0.000980 \\
$\textbf{RBE}^{(Nor)}_{\alpha{=}0.3}$&0.000345 &-0.001696&  0.000271&  \textbf{0.000153}&  0.000569\\
$\textbf{RBE}^{(Nor)}_{\alpha{=}0.9}$&0.000040& -0.001952&  0.000287&  0.000161 & \textbf{0.000226} \\
$\textbf{RBE}^{(Dir)}_{\alpha{=}0.3}$&0.000298& -0.000962&  0.000857&  0.000380&  0.000592\\
$\textbf{RBE}^{(Dir)}_{\alpha{=}0.9}$&0.000254& \textbf{-0.000911}&  \textbf{0.000096}&  0.000379& 0.000418
 \\
$\textbf{RBE}^{(Ord)}_{\alpha{=}0.3}$&-0.000078& -0.004988&  0.000496& -0.000193&  0.000510\\
$\textbf{RBE}^{(Ord)}_{\alpha{=}0.9}$& \textbf{-0.000022}& -0.004948&  0.000479& -0.000659&  0.000518\\
\botrule
\end{tabular}
\end{table}

\begin{table}[htb!]
\caption{RMSE of the estimates with Gompertz lifetime distribution.}\label{tab3b}%
\setlength{\tabcolsep}{2.5pt} 
\begin{tabular}{@{}lllllllllll@{}}
\toprule
& \multicolumn{5}{@{}c@{}}{\textbf{Pure data}}\\\cmidrule(lr){2-6}
&$\bm{\hat{c}}_0$&$\bm{\hat{c}}_1$&$\bm{\hat{\gamma}}_1$&$\bm{\hat{\gamma}}_2$&$\bm{\hat{\gamma}}_3$\\
\midrule
$\textbf{MLE}$
& 0.004964 & 0.022120 & 0.002859 & 0.004344 & 0.001554 \\ 
$\textbf{MDE}_{\alpha{=}0.3}$&0.008213& 0.018336& 0.004255& 0.003567& 0.001730
 \\
$\textbf{MDE}_{\alpha{=}0.9}$
& 0.009188 & 0.015972 & 0.005002 & 0.007685 & \textbf{0.000274} \\
$\textbf{BE}^{(Nor)}$
& 0.001768 & 0.001390 & \textbf{0.001029} & 0.001304 & 0.001058 \\
$\textbf{BE}^{(Dir)}$
& 0.001786 & 0.001310 & \textbf{0.001020} & 0.001275 & 0.001149 \\
$\textbf{BE}^{(Ord)}$
& \textbf{0.001735} & \textbf{0.001285} & 0.001075 & \textbf{0.001221} & 0.001035 \\
$\textbf{RBE}^{(Nor)}_{\alpha{=}0.3}$&0.002282& 0.001756 &0.001037 &0.001705& 0.000934
 \\
$\textbf{RBE}^{(Nor)}_{\alpha{=}0.9}$
& 0.001939 & 0.001589 & 0.001337 & 0.001514 & 0.001445 \\
$\textbf{RBE}^{(Dir)}_{\alpha{=}0.3}$&0.002295& 0.001737&0.001007& 0.001730& 0.001391
 \\
$\textbf{RBE}^{(Dir)}_{\alpha{=}0.9}$
& 0.001845 & 0.001593 & 0.001335 & 0.001481 & 0.001432 \\
$\textbf{RBE}^{(Ord)}_{\alpha{=}0.3}$&0.002040& 0.005187 &0.001413& 0.009663& 0.001379
 \\
$\textbf{RBE}^{(Ord)}_{\alpha{=}0.9}$
& 0.001945 & 0.001610 & 0.001329 & 0.001565 & 0.001351 \\
\midrule
& \multicolumn{5}{@{}c@{}}{\textbf{Contamination}}\\\cmidrule(lr){2-6}
&$\bm{\tilde{c}}_0$&$\bm{\tilde{c}}_1$&$\bm{\tilde{\gamma}}_1$&$\bm{\tilde{\gamma}}_2$&$\bm{\tilde{\gamma}}_3$\\
\midrule
$\textbf{MLE}$&0.013473& 0.040874& 0.032345& 0.014718&0.003576
 \\ 
$\textbf{MDE}_{\alpha{=}0.3}$&0.010553& 0.023214& 0.006442& 0.009742& 0.000526
\\
$\textbf{MDE}_{\alpha{=}0.9}$&0.009328& 0.017156 &0.005144& 0.008158 &0.000271
 \\
$\textbf{BE}^{(Nor)}$&0.005688& 0.010078& 0.004786 &0.006672& 0.002565
\\
$\textbf{BE}^{(Dir)}$&0.006860 &0.010019& 0.004995 &0.006378& 0.002753
 \\
$\textbf{BE}^{(Ord)}$&0.004939& 0.010097& 0.004530& 0.007007& 0.002001
 \\
$\textbf{RBE}^{(Nor)}_{\alpha{=}0.3}$&\textbf{0.001883}& 0.001643 &\textbf{0.001333}& 0.001562& \textbf{0.001371}
\\
$\textbf{RBE}^{(Nor)}_{\alpha{=}0.9}$&0.002036& 0.001623 &0.001531& 0.001667 &0.001439
\\
$\textbf{RBE}^{(Dir)}_{\alpha{=}0.3}$&0.001951& \textbf{0.001575} &0.001428& \textbf{0.001520}& 0.001391
 \\
$\textbf{RBE}^{(Dir)}_{\alpha{=}0.9}$&0.001938 &0.001695 &0.001523& 0.001689& 0.001482
 \\
$\textbf{RBE}^{(Ord)}_{\alpha{=}0.3}$&0.002047 &0.005141 &0.001376& 0.009470& 0.001453
 \\
$\textbf{RBE}^{(Ord)}_{\alpha{=}0.9}$&0.002052& 0.005109& 0.001578& 0.009801& 0.001569
\\
\botrule
\end{tabular}
\end{table}

\begin{figure}[htb!]
\begin{center}
\subfloat[Absolute Bias (Weibull)]{\includegraphics[height=5.5cm,width =0.5\textwidth]{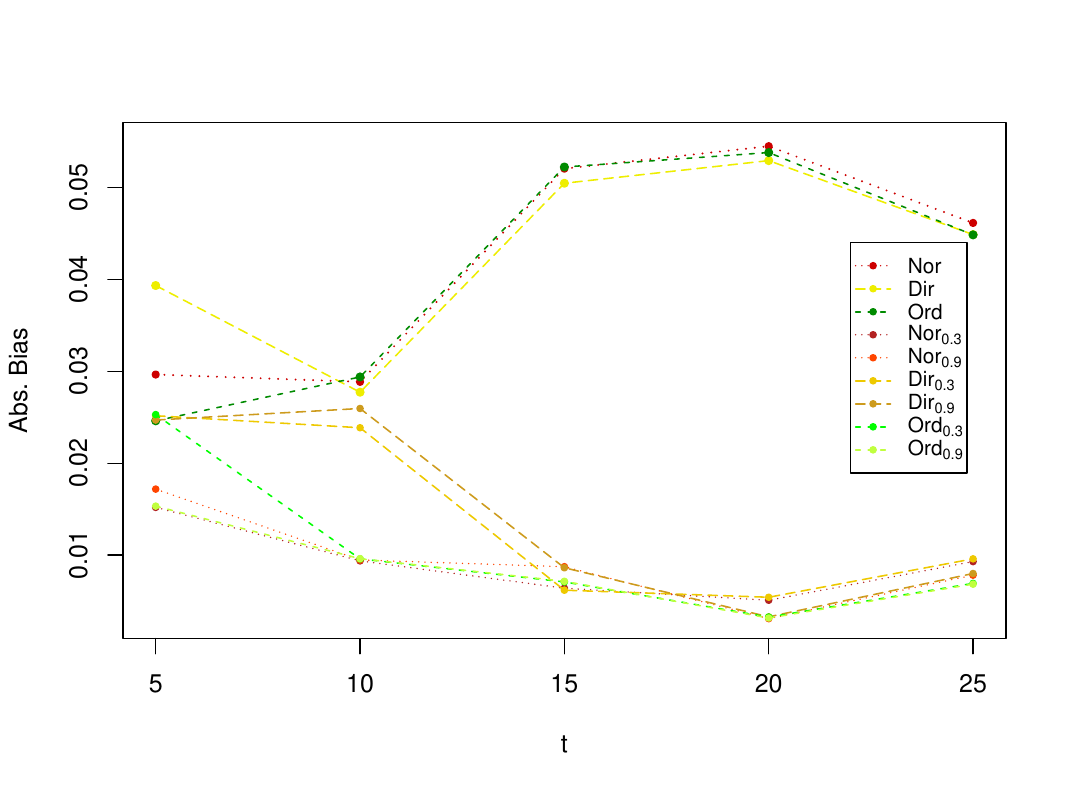}} 
\subfloat[MSE (Weibull)]{\includegraphics[height=5.5cm,width =0.5\textwidth]{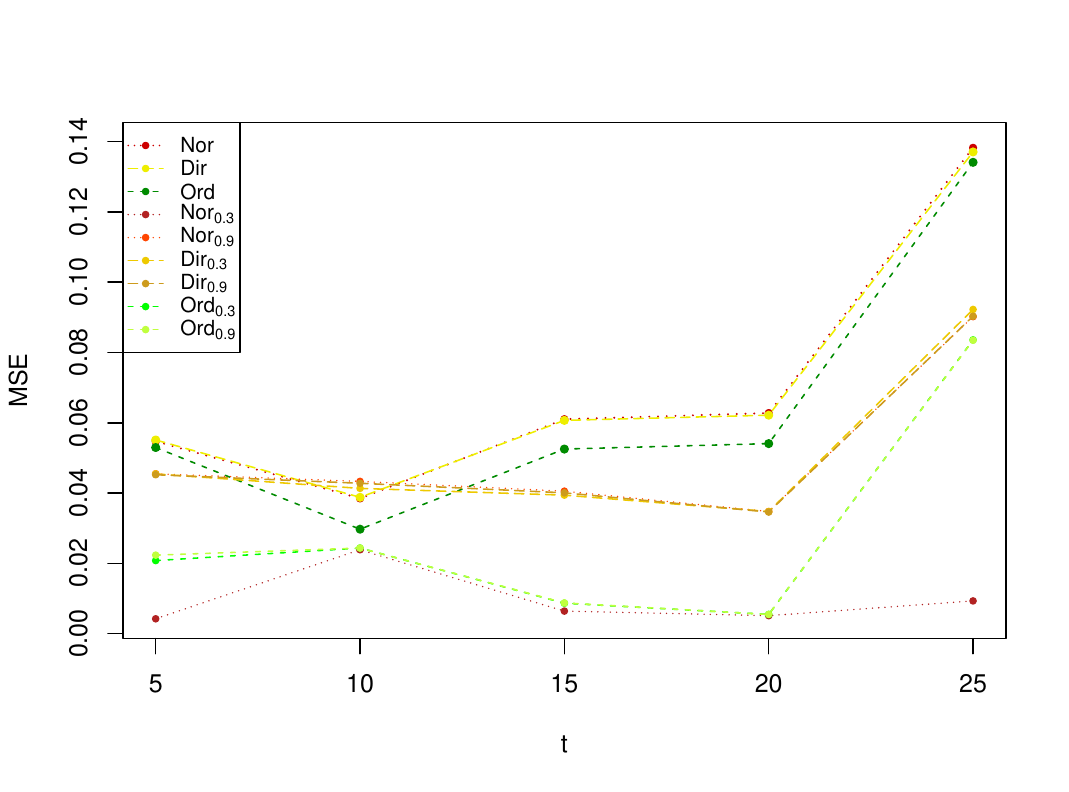}}\\
\subfloat[Absolute Bias (Gompertz)]{\includegraphics[height=5.5cm,width =0.5\textwidth]{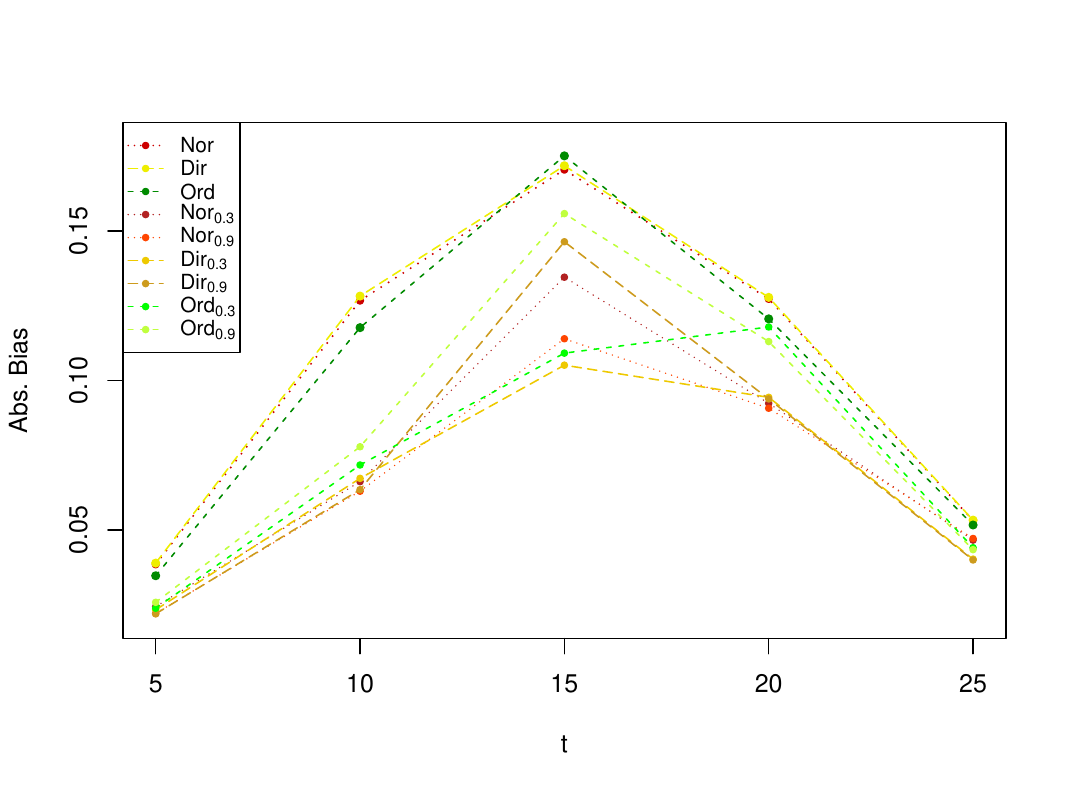}} 
\subfloat[MSE (Gompertz)]{\includegraphics[height=5.5cm,width =0.5\textwidth]{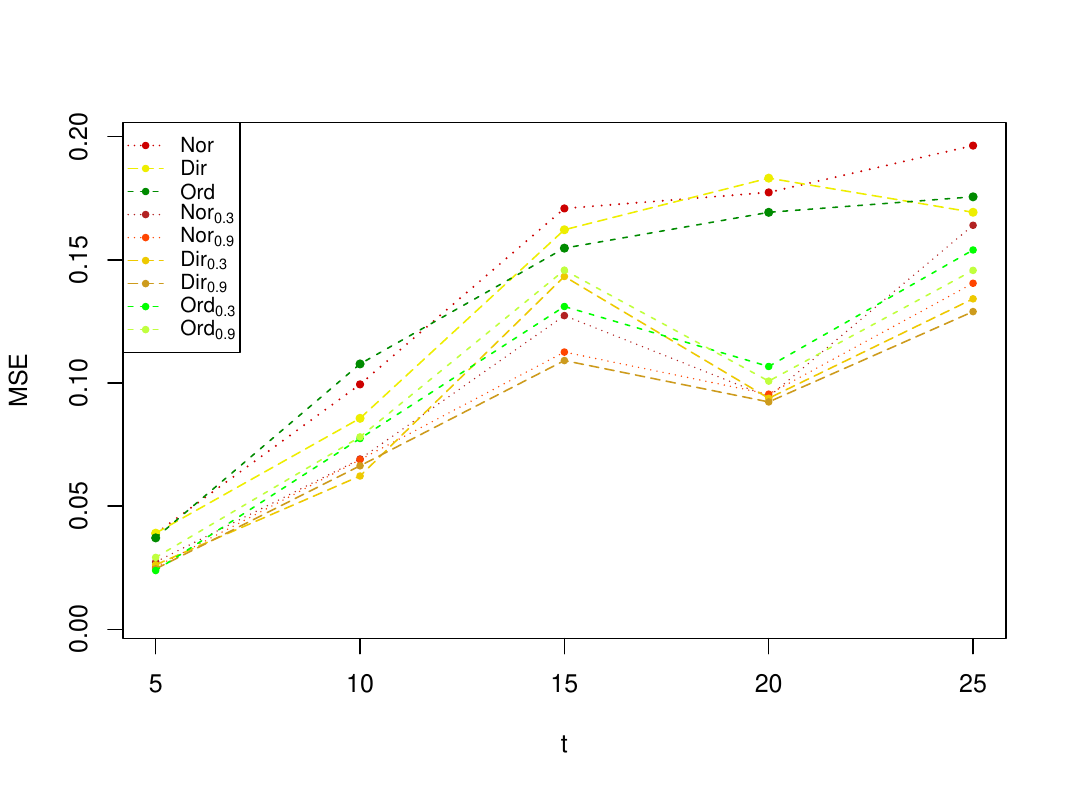}}\\
\end{center}
\caption{Absolute bias and MSE for reliability estimates with Weibull and Gompertz lifetime distribution under contamination .}
\label{bzzm}
\end{figure}

Robustness can be observed through the bias of the estimates.  Hence, bias and root mean square error (RMSE) are obtained through Monte Carlo simulation based on 1000 generations.  The maximum likelihood estimate (MLE) and minimum density power divergence estimate (MDE) have also been obtained for comparative purposes.  The coordinate descent method \cite{baghel2024analysis,baghel2024robust} is used to derive MLE and MDE.  The Bayes estimate (BE) and robust Bayes estimate (RBE) are obtained by using Hamiltonian Monte Carlo (HMC) given in the algorithm \ref{alg}.  Three chains of $N=1200$ values are generated through HMC, and the first $N^{'}=200$ values from each chain are discarded as burn-in period.  For ordered restricted prior, we set hyperparameter $\mu_j$ as the true value of $c_j$ and $\sigma_j^2$ as the variance of $c_j.$  The asymptotic variance-covariance matrix for the parameters $\theta$ is in the appendix.  For the other hyperparameters when we set $b_0=(0.4,1)$ and $a_1=(0.4,1)$, we obtain $a_0=(\sum_{i=1}^{3}\gamma_i)b_0$, $a_2=\frac{\gamma_2}{\gamma_1}a_1$ and $a_3=\frac{\gamma_3}{\gamma_1}a_1$; for Weibull and Gompertz lifetime distributions respectively.  The outcomes are reported in tables \eqref{tab2a}-\eqref{tab3b}.  The smallest magnitude of bias and smallest RMSE for each of the parameters in pure and contaminated schemes are highlighted in bold.  A lower magnitude of bias and RMSE indicate better accuracy and precision, respectively.

Table \eqref{tab2a} compares the bias of estimators under the Weibull lifetime distribution in pure and contaminated data settings.  In pure data settings, MLE exhibits a lower magnitude of bias than MDE across all parameters, while BE generally outperforms RBE.  Hence, BE performs well without contamination, though RBE offers bias values closer but slightly lower magnitude than BE.  However, the advantage of RBE becomes more pronounced when contamination is present, where MLE suffers the most, displaying the highest magnitude of biases.  The MDE showcase lower absolute bias than MLE, with its performance depending on tuning parameter $\alpha.$  Here $\alpha=0.3$ tends to introduce more bias.  Though BE is superior to MLE and MDE, it is more sensitive to contamination than RBE.  RBE performs best under contamination as the magnitude of bias remains consistently smaller than the other three estimators.  The RBE under Dirichlet and order restricted prior demonstrate the lowest magnitude of bias across parameters.  It is also observed that increase in bias from pure to contamination in MDE and RBE is comparatively lower than that of MLE and BE, proving the robustness of DPD based estimation methods. 
 Table \eqref{tab2b} presents RMSE of estimators under a Weibull lifetime distribution for pure and contaminated data settings.  It is observed that Bayesian methods (BE and RBE) demonstrate superior performance over classical methods (MLE and MDE) in both pure and contaminated settings as they exhibit lower RMSE.  While BE offers lower RMSE under pure data, it is more affected by contamination than RBE.  Overall, RBE emerges as the best-performing method with order-restricted prior in the presence of contamination, achieving the lowest magnitude of bias and RMSE in general and making it the most precise and robust choice in the current setup.

Tables \eqref{tab3a} and \eqref{tab3b} demonstrate bias and RMSE of estimators under the Gompertz lifetime distribution for pure and contaminated datasets.  It can be observed that MLE is highly sensitive to contamination, leading to highest bias and RMSE compared to other methods of estimation.  MDE holds robustness but still has a relatively higher magnitude of bias and RMSE than Bayesian methods. 
 The tables indicate that while BE is satisfactory in pure data scenarios, RBE is more effective under contamination, as it exhibits less magnitude of bias and RMSE than BE.  It is also observed that an increase in bias from pure to contaminated data is lower for RBE than for BE.  Thus, RBE can be referred to as a robust estimation method.  The classical estimates (MLE and MDE) underperform relative to Bayes estimates (BE and RBE), which is reasonably expected.  For RBE under contamination, no prior clearly demonstrates overall superiority in terms of bias in table \eqref{tab3a}.  However, RBE with normal prior and Dirichlet prior attains the lowest RMSE under contamination in table \eqref{tab3b}.  The prior selection depends on the priority given to robustness or precision in real-life situations. 
 Further, the absolute bias (Abs.Bias) and MSE of reliability estimates for parameters in the contaminated settings are plotted in Figure \ref{bzzm} to illustrate robustness graphically.  The better performance of RBE over BE under contamination is also visible from these figures.

\subsection{\textbf{Optimal choice of tuning parameter}}\label{optune}
As discussed in the introduction, the DPD measure-based estimation depends on the choice of tuning parameter $\alpha$.  Hence, finding the optimal value for tuning the parameter concerning the interest criteria is required \cite{bas2021,cas2022,sugasawa2021selection,yonekura2023adaptation}. Here, We suggest a non-iterative method based on the approach introduced by Warwick and Jones \cite{w2005}, which involves minimizing the objective function
\begin{equation}
\Phi_{\alpha}(\hat{\bm{\theta}})=C_1\,D_{\alpha}({\hat{\bm{\theta}}})+C_2\,tr\Big(J^{-1}_{\alpha}({\hat{\bm{\theta}}})K_{\alpha}({\hat{\bm{\theta}}})J^{-1}_{\alpha}({\hat{\bm{\theta}}})\Big), \label{tune}
\end{equation}
where $D_{\alpha}(\hat{\bm{\theta}})$ is the DPD measure indicating robustness, $tr(.)$ is a trace of variance-covariance matrix handling precision of the estimates and $C_1, C_2$ are predefined positive weight values with $C_1+C_2=1$.  Unlike Warwick and Jones \cite{w2005} and Basak et al. \cite{bas2021} methods, this approach doesn't need any pilot estimator.  The expression for the variance-covariance matrix of the estimates is given in the appendix.  The values of $C_1$ and $C_2$ are situation-specific, whether more weight should be given to robustness or precision.  When the dataset contains heavy outliers, maintaining high precision becomes difficult.  In this situation, allowing more weight to robustness is quite reasonable.  In this study, we have given equal weight values $C_1=C_2=0.5$ for the data analysis.

\section{Data analysis}\label{sec7}
For practical implementation of the results previously discussed, a dataset examining the reliability of light bulbs is taken here from the experimental study conducted by Zhu \cite{zhu2010optimal}.  Balakrishnan et al. \cite{balakrishnan2023robust} used this data for robust inference of nondestructive one-shot device testing data under step-stress ALT.  In light bulb experiment $n=64$, miniature light bulbs are put into a two-step SSALT experiment where voltage $x_1=2.25V$ is applied for up to $\tau_1=96$ hours, and then the voltage is increased to $x_2=2.44V$.  The stopping time of the experiment is $\tau_2=140$ hours.  The failure times of light bulbs within 140 hours are recorded as follows.\\

\noindent 12.07, 19.5, 22.1, 23.11, 24, 25.1, 26.9, 36.64, 44.1, 46.3, 54, 58.09, 64.17, 72.25, 86.9, 90.09, 91.22, 102.1, 105.1, 109.2, 114.4, 117.9, 121.9, 122.5, 123.6, 126.5, 130.1, 14, 17.95, 24, 26.46, 26.58, 28.06, 34, 36.13, 40.85, 41.11, 42.63, 52.51, 62.68, 73.13, 83.63, 91.56, 94.38, 97.71, 101.53, 105.11, 112.11, 119.58, 120.2, 126.95, 129.25, 136.31.\\

$n_s=11$ lightbulbs survived after the termination of the experiment.  Failure times are multiplied by 0.2 and 0.1 for Weibull and Gompertz lifetime distributions for computational convenience, respectively.  Intermediate inspection times on scaled failure times are then taken as $\tau_1{=}(6.4, 12.8, 19.2, 22.2, 25.2, 28.0)$ and $\tau_2{=}(3.2, 6.4, 9.6, 11.1, 12.6, 14.0)$ for Weibull and Gompertz lifetime distributions, respectively.  To test if the standard Lehman family of distribution is fitted to data for the given model, a bootstrap-based goodness of fit test is performed, and an approximated p-value is obtained.  The distance-based test statistic to conduct testing is defined as
\begin{equation}
    TS=\left\vert\frac{n_s-\hat{n}_s}{\hat{n}_s}\right\vert+\sum_{i=1}^{k}\sum_{m=1}^{q_i}\left\vert\frac{n_{im}-\hat{n}_{im}}{\hat{n}_{im}}\right\vert.\label{test}
\end{equation}
Here, $\hat{n}_{im}$ and $\hat{n}_s$ are the estimates of the expected number of failures and survivals obtained through MLE.  Since the latency period was not observed by Zhu \cite{zhu2010optimal} in her experiment, here the lag period $\delta$ is unknown.  To find an estimate of lag period $\delta$, an extensive grid search is performed as discussed in Section \ref{2.2.2}. 
 The value of $\delta$ for which the likelihood function is maximized is found here as $\hat{\delta}=0.001.$  For deriving BE and RBE, HMC is used where we consider $\epsilon=0.001$, $L=10$, $v=(0.01,0.01,0.01,0.01)$ and $M=1/v.$  For Dirichlet prior $\sigma^2_{(p)}=0.05$ is taken.
\begin{table}[htb!]
\caption{Information regarding the analysis under special cases from the Lehman family of distributions for the data.}\label{tab6}
\begin{tabular}{@{}lll@{}}
\toprule
\textbf{Information}&\textbf{Weibull}&\textbf{Gompertz}\\
\midrule
\textbf{Test statistic}& $TS_1=8.483801$ &$TS_2=4.013871$\\
\textbf{p-value}& $0.660$& $0.836$\\
\botrule
\end{tabular}
\end{table}

\begin{table}[htb!]
\caption{Parameter estimates (95\% Asymp.CI\slash HPD CRI) with Weibull lifetime distribution for the data.}\label{tab8}%
\begin{tabular}{@{}lcccc@{}}
\toprule
& {$\bm{\hat{c}}_0$}&{$\bm{\hat{c}}_1$}&{$\bm{\hat{\gamma}}_1$}&{$\bm{\hat{\gamma}}_2$}\\ 
\cmidrule(lr){2-5}
& \textbf{Est.(CI)}&\textbf{Est.(CI)}& \textbf{Est.(CI)}&\textbf{Est.(CI)}\\
\midrule
$\textbf{MLE}$&  -0.900837 & 0.060338 & 0.200091 & 0.699624 \\
& (-0.94048,-0.86119) &(0.01860,0.10207) &(0.11944,0.28074) &(0.67893,0.72031) \\
$\textbf{MDE}$&  -0.899641 & 0.060466 & 0.199989 & 0.699673 \\
&  (-0.95477,-0.84451)&(0.00594,0.11498)&(0.10746,0.29250)&(0.67630,0.72304)\\
\midrule
& \textbf{Est.(HPD CRI)}&\textbf{Est.(HPD CRI)}& \textbf{Est.(HPD CRI)}&\textbf{Est.(HPD CRI)}\\
\midrule
$\textbf{BE}^{(Nor)}$&  -0.901031 & 0.060688 & 0.200133 & 0.701713 \\
& (-0.90393,-0.89816) &(0.05762,0.06350) &(0.19740,0.20299) &(0.69898,0.70463) \\
$\textbf{BE}^{(Dir)}$&  -0.899570 & 0.059718&  0.201738& 0.700483  \\
& (-0.90270,-0.89653) & (0.05668,0.06282)&(0.19851,0.20458) &(0.69718,0.70341) \\
$\textbf{BE}^{(Ord)}$&  -0.910319 & 0.069433 & 0.219312 & 0.708060  \\
&(-0.91312,-0.90711)  & (0.06629,0.07213)&(0.21618,0.22210) & (0.70517,0.71149) \\
$\textbf{RBE}^{(Nor)}_{\alpha{=}0.2}$&  -0.898346&  0.060128 & 0.198984 & 0.698524  \\
& (-0.90137,-0.89501) &(0.05738,0.06325) &(0.19561,0.20193) &(0.69557,0.70149) \\
$\textbf{RBE}^{(Nor)}_{\alpha{=}0.9}$& -0.899722 & 0.059221 & 0.201124 & 0.700870 \\
& (-0.90283,-0.89673) &(0.05618,0.06246) &(0.19823,0.20423) & (0.69788,0.70401)\\
$\textbf{RBE}^{(Dir)}_{\alpha{=}0.2}$& -0.901835&  0.059078&  0.198028 & 0.698202  \\
& (-0.90270,-0.89653) & (0.05668,0.06282)&(0.19851,0.20458) & (0.69718,0.70341)\\
$\textbf{RBE}^{(Dir)}_{\alpha{=}0.9}$&  -0.899526 & 0.060044 & 0.199355 & 0.699796 \\
&(-0.90247,-0.89666)  & (0.05727,0.06331)&(0.19629,0.20251) & (0.69675,0.70265)\\
$\textbf{RBE}^{(Ord)}_{\alpha{=}0.2}$&  -0.898733& 0.060684&  0.201255 & 0.699638 \\
& (-0.90163,-0.89556) & (0.05813,0.06389) &(0.19836,0.20422) & (0.69648,0.70305) \\
$\textbf{RBE}^{(Ord)}_{\alpha{=}0.9}$&  -0.899588 & 0.061308 & 0.199218 & 0.701023 \\
& (-0.90271,-0.89683) &(0.05809,0.06402) &(0.19617,0.20218) &(0.69703,0.70346)  \\
\botrule
\end{tabular}
\end{table}

\begin{table}[htb!]
\caption{Parameter estimates (95\% Asymp.CI\slash HPD CRI) with Gompertz lifetime distribution for the data.}\label{tab9}%
\begin{tabular}{@{}lcccc@{}}
\toprule
& {$\bm{\hat{c}}_0$}&{$\bm{\hat{c}}_1$}&{$\bm{\hat{\gamma}}_1$}&{$\bm{\hat{\gamma}}_2$}\\ 
\cmidrule(lr){2-5}
& \textbf{Est.(CI)}&\textbf{Est.(CI)}& \textbf{Est.(CI)}&\textbf{Est.(CI)}\\
\midrule
$\textbf{MLE}$&-0.200276 & 0.219982 & 0.070031 & 0.085010\\
&(-0.27680,-0.12374)& (0.03555,0.40441)& (0.03242,0.10763)&(0.06137,0.10865)\\
$\textbf{MDE}$&-0.200304 & 0.219993 & 0.069999& 0.085004 \\
& (-0.29214,-0.10846) &  (0.06660,0.37338)&  (0.04392,0.09607)& (0.06802,0.10198)\\
\midrule
& \textbf{Est.(HPD CRI)}&\textbf{Est.(HPD CRI)}& \textbf{Est.(HPD CRI)}&\textbf{Est.(HPD CRI)}\\
\midrule
$\textbf{BE}^{(Nor)}$& -0.198350 & 0.219059 & 0.071431 & 0.085721\\
&(-0.20137,-0.19524)
 &(0.21610,0.22207) &(0.06860,0.07431) &(0.08300,0.08877)\\
$\textbf{BE}^{(Dir)}$& -0.200100 & 0.219229 & 0.069270 & 0.087004
  \\
&(-0.20301,-0.19684) &(0.21610,0.22255 &(0.06620,0.07194) &(0.08356,0.08992)\\
$\textbf{BE}^{(Ord)}$&  -0.201832 & 0.219690 & 0.071144&  0.083926 \\
&(-0.20527,-0.19870)  &(0.21679,0.22314) &(0.06821,0.07439)& (0.08079,0.08724) \\
$\textbf{RBE}^{(Nor)}_{\alpha{=}0.2}$& -0.201931&  0.218025 & 0.070149 & 0.084411  \\
&(-0.20495,-0.19876) &(0.21491,0.22099) &(0.06686,0.07330) &(0.08123,0.08721)\\
$\textbf{RBE}^{(Nor)}_{\alpha{=}0.9}$& -0.198917&0.220460&0.070556&  0.085723 \\
&(-0.20210,-0.19623) &(0.21723,0.22354) & (0.06768,0.07375)&(0.08269,0.08883) \\
$\textbf{RBE}^{(Dir)}_{\alpha{=}0.2}$&-0.201619 & 0.221955 & 0.069912 & 0.084039 \\
&(-0.20507,-0.19900) &(0.21892,0.22485) &(0.06688,0.07285) &(0.08076,0.08703)\\
$\textbf{RBE}^{(Dir)}_{\alpha{=}0.9}$& -0.200165 & 0.219366  &0.071329 & 0.084549\\
& (-0.20354,-0.19717)&(0.21614,0.22222) &(0.06815,0.07436) &(0.08159,0.08786) \\
$\textbf{RBE}^{(Ord)}_{\alpha{=}0.2}$&  -0.198714 & 0.220349 & 0.070930 & 0.086112 \\
&(-0.20151,-0.19564)  &(0.21719,0.22344) &(0.06784,0.07409)& (0.08290,0.08930) \\
$\textbf{RBE}^{(Ord)}_{\alpha{=}0.9}$& -0.198691 & 0.221382&  0.068357 & 0.084250  \\
&(-0.20194,-0.19624)  & (0.21830,0.22413)
&(0.06533,0.07133) &  (0.08096,0.08738) \\
\botrule
\end{tabular}
\end{table}

A bootstrap-based goodness of fit test is conducted with the test statistic given in the equation \eqref{test}.  The value of test statistics and corresponding p-value for both lifetime distributions are reported in table \ref{tab6}.  The significant p-values indicate the suitability of both lifetime distributions to the data.

The estimates derived from BE and RBE with $95\%$ highest posterior density credible interval ((HPD CRI) along with MLE and MDE with $95\%$ asymptotic confidence intervals (CI) are reported in tables \ref{tab8} and \ref{tab9}.  It is found that the optimal value of the tuning parameter found through equation \eqref{tune} is $\alpha_{opt}=(0.60, 0.70)$ under Weibull and Gompertz lifetime assumptions, respectively, for the lightbulb data.  The MDE estimates are obtained at $\alpha_{opt}$ values.  The bootstrap bias (BT Bias) and bootstrap root mean square of error (BT RMSE) of the estimates are given in table \ref{boot}.  From this table, we observe a smaller magnitude of BT Bias and BT RMSE for RBE compared to other estimation methods.

\begin{table}[htb!]
\caption{BT Bias and BT RMSE of the estimates.}\label{boot}%
\setlength{\tabcolsep}{2.6pt} 
\begin{tabular}{@{}lllllllll@{}}
\toprule
& \multicolumn{4}{@{}c@{}}{\textbf{BT Bias}}
& \multicolumn{4}{@{}c@{}}{\textbf{BT RMSE}}\\\cmidrule(lr){2-5}\cmidrule(lr){6-9}
&$\bm{\hat{c}}_0$&$\bm{\hat{c}}_1$&$\bm{\hat{\gamma}}_1$&$\bm{\hat{\gamma}}_2$&$\bm{\hat{c}}_0$&$\bm{\hat{c}}_1$&$\bm{\hat{\gamma}}_1$&$\bm{\hat{\gamma}}_2$\\
\midrule
& \multicolumn{8}{@{}c@{}}{\textbf{Weibull lifetime distribution}}\\
\midrule
$\textbf{MLE}$&0.004267 & 0.006088& -0.018654& -0.016093&0.004828& 0.006648 &0.019801 &0.016239 \\
$\textbf{MDE}$&0.000291&  0.001724& -0.006046& -0.005493&\textbf{0.000711}& 0.001924& 0.006442 &0.005526 \\
$\textbf{BE}^{(Nor)}$&-0.000193 & 0.000121&  0.000279 &  0.000095 &0.001163& 0.001179& 0.001168 &0.001180 \\
$\textbf{BE}^{(Dir)}$&-0.000196 &-0.000275&-0.000091&  0.000239 &0.001149& 0.001167& 0.001151&0.001150 \\
$\textbf{BE}^{(Ord)}$&   0.000846& -0.000425& -0.000100&  0.000306 &0.001423& 0.001240 &0.001200& 0.001199  \\
$\textbf{RBE}^{(Nor)}_{\alpha{=}0.2}$&\textbf{0.000011} &-0.000047 &-0.000072 & 0.000026 &0.001180& 0.001134& \textbf{0.001125} &\textbf{0.001141}\\
$\textbf{RBE}^{(Nor)}_{\alpha{=}0.9}$& 0.000017&  0.000029 &-0.000106 & 0.000071 &0.001200& 0.001184& 0.001159& 0.001190  \\
$\textbf{RBE}^{(Dir)}_{\alpha{=}0.2}$& -0.000025&  0.000079& \textbf{0.000039}& -0.000018&0.001138& 0.001159& 0.001150 &0.001155\\
$\textbf{RBE}^{(Dir)}_{\alpha{=}0.9}$&  0.000054& -0.000582& -0.000056& 0.000316&0.001166& 0.001307& 0.001188& 0.001218 \\
$\textbf{RBE}^{(Ord)}_{\alpha{=}0.2}$&0.000037 & \textbf{0.000007}& -0.000097&  \textbf{0.000001}&0.001122&0.001140 &0.001148& 0.001155\\
$\textbf{RBE}^{(Ord)}_{\alpha{=}0.9}$&  0.000387& -0.000174& 0.000063& 0.000264 &0.001204& \textbf{0.001131}& 0.001142 &0.001205 \\
\midrule
& \multicolumn{8}{@{}c@{}}{\textbf{Gompertz lifetime distribution}}\\
\midrule
$\textbf{MLE}$& -0.001689& -0.005876& -0.025580& -0.047481&0.017452& 0.063652& 0.053224& 0.070543\\
$\textbf{MDE}$&-0.000312& -0.000320& -0.009659& -0.030562&\textbf{0.000874}& 0.001261& 0.012133& 0.037575 \\
$\textbf{BE}^{(Nor)}$&-0.000094&  0.000105 & 0.000218 & 0.000097&0.001191& 0.001137& 0.001173&0.001201 \\
$\textbf{BE}^{(Dir)}$&0.000103& -0.000395 & 0.000112 &-0.000166&0.001173& 0.001160& 0.001142& 0.001124
\\
$\textbf{BE}^{(Ord)}$&0.000102 & 0.000076& -0.000074 &-0.000105 &0.001157& 0.001155& 0.001148 &0.001153 \\
$\textbf{RBE}^{(Nor)}_{\alpha{=}0.2}$&0.000015 & 0.000086 & \textbf{0.000011}& \textbf{-0.000012}&0.001130& 0.001158 &0.001126 &0.001179 \\
$\textbf{RBE}^{(Nor)}_{\alpha{=}0.9}$&0.000049 & \textbf{0.000008}& -0.000023 & 0.000054&0.001172& 0.001173& 0.001185& \textbf{0.001105} \\
$\textbf{RBE}^{(Dir)}_{\alpha{=}0.2}$&-0.000022& -0.000015& -0.000063 &-0.000047&0.001189& \textbf{0.001125}& 0.001147& 0.001191\\
$\textbf{RBE}^{(Dir)}_{\alpha{=}0.9}$& -0.000153& -0.000029 &-0.000074& -0.000050&0.001175& 0.001140 &0.001208& 0.001174 \\
$\textbf{RBE}^{(Ord)}_{\alpha{=}0.2}$& -0.000032 &-0.000040&-0.000017& -0.000156&0.001159& 0.001135& \textbf{0.001124}& 0.001154\\
$\textbf{RBE}^{(Ord)}_{\alpha{=}0.9}$&  \textbf{0.000001}&  0.000013&  0.000055& -0.000053&0.001152& 0.001176 &0.001169 &0.001161
 \\
\botrule
\end{tabular}

\end{table}

\subsection{\textbf{Testing of hypothesis based on robust Bayes factor}}
For testing of hypothesis, the robust Bayes factor is used as a test statistic for any particular hypothesis for the given data.  Let us define a simple null hypothesis against an alternative hypothesis as
$$ 	\bm{H}_0 : \bm{\theta}=\bm{\theta}_0\quad \text{against}\quad \bm{H}_1 : \bm{\theta}\neq \bm{\theta}_0.$$
A continuous prior density would lead to zero prior probability to test $\bm{H}_0$.  Therefore, it is suggestive to take an $\varepsilon$-neighborhood (spherical) around $\bm{\theta}_0$ and assign prior probability $\rho_0$ under $\bm{H}_0$.  The empirical prior and posterior probabilities are calculated to obtain the empirical Bayes factor.  From equation \eqref{odds}, the Bayes factor can be calculated using relation $$\text{Posterior odds}=\text{Prior odds}\times \text{Bayes factor}.$$

Here, the simple null hypothesis under Weibull lifetime distribution to be tested is $\bm{\theta}^{(1)}_0=(-0.09,0.06,0.2,0.7)^{'}$ and $\varepsilon=0.003.$  The values of empirical Bayes factor (BF$_{01}$) are reported in table \ref{tab13}.  The interpretation of Bayes factor values (BF$_{01}$) can be made based on the scale given in table \ref{tab1}.  Since the BF$_{01}$ value lies in $20$ to $150$ under all three priors, support for $\bm{H}_0$ is strong.  For Gompertz lifetime distribution $\bm{\theta}^{(2)}_0=(-0.2,0.22,0.07,0.085)^{'}$  and $\varepsilon=0.0028$ is taken. Bayes factor values are provided in table \ref{tab14}.  From the interpretation given in table \ref{tab1}, support for $\bm{H}_0$ under Normal prior is positive ($\alpha=0.7,0.9$) and strong ($\alpha=0.2$), under Dirichlet prior is positive and under order-restricted prior, support for $\bm{H}_0$ is positive ($\alpha=0.2$) and strong ($\alpha=0.7,0.9$).

\begin{table}[htb!]
\caption{Empirical value of Bayes factor with Weibull lifetime distribution.}\label{tab13}%
\begin{tabular}{@{}llll@{}}
\toprule
\textbf{Tuning}	& \textbf{Prior}&\textbf{Posterior}&\textbf{Bayes Factor} \\
\textbf{Parameter}&\textbf{odds}&\textbf{odds}& \textbf{BF$_{01}$} \\
\midrule
\multicolumn{4}{c}{\textbf{Normal prior}}\\
\midrule
$\bm{0.2}$&\multirow{3}*{0.216463}&11.625000&53.704328 \\
$\bm{0.6}$& & 19.199980&88.698669 \\
$\bm{0.9}$& &11.875436  & 54.861274 \\
\midrule
 \multicolumn{4}{c}{\textbf{Dirichlet prior}}\\
 \midrule
 $\bm{0.2}$&\multirow{3}*{0.342281}&19.600000 &57.262892  \\
$\bm{0.6}$& &15.666670 & 45.771456 \\
$\bm{0.9}$& &10.763211 & 31.445540\\
\midrule
 \multicolumn{4}{c}{\textbf{Order restriction}}\\
 \midrule
$\bm{0.2}$&\multirow{3}*{0.342281}&16.166670 &47.232157  \\
$\bm{0.6}$& &11.875000 & 34.693716 \\
$\bm{0.9}$& &16.166670 &47.232157 \\
\botrule
\end{tabular}
\end{table}

\begin{table}[htb!]
\caption{Empirical value of Bayes factor with Gompertz lifetime distribution.}\label{tab14}%
\begin{tabular}{@{}llll@{}}
\toprule
\textbf{Tuning}	& \textbf{Prior}&\textbf{Posterior}&\textbf{Bayes Factor} \\
\textbf{Parameter}&\textbf{odds}&\textbf{odds}& \textbf{BF$_{01}$} \\
\midrule
\multicolumn{4}{c}{\textbf{Normal prior}}\\
\midrule
$\bm{0.2}$&\multirow{3}*{0.274760}&5.968748 & 21.723496 \\
$\bm{0.7}$& &5.058825 & 18.411795\\
$\bm{0.9}$& &5.151514  &  18.749141\\
\midrule
 \multicolumn{4}{c}{\textbf{Dirichlet prior}}\\
 \midrule
$\bm{0.2}$&\multirow{3}*{0.351351}&6.692308 & 19.047357\\
$\bm{0.7}$& &5.028571 & 14.312101\\
$\bm{0.9}$& &5.393940&15.351998 \\
\midrule
 \multicolumn{4}{c}{\textbf{Order restriction}}\\
 \midrule
$\bm{0.2}$&\multirow{3}*{ 0.328947}& 5.571429& 16.937163 \\
$\bm{0.7}$& &6.615386 & 20.110795 \\
$\bm{0.9}$& &8.699998 & 26.448023\\
\botrule
\end{tabular}
\end{table}

\section{Conclusion} \label{sec8}
The present study has incorporated the cumulative risk model to determine the lifetime prognosis of the nondestructive one-shot device (NOSD) under a step-stress accelerated life testing experiment where the lifetime of NOSD comes from a standard Lehman family of distributions.  The robust estimation procedure has been developed in the Bayesian framework, where the robustified posterior involved an exponential form of maximiser equation based on density power divergence.  The Hamiltonian Monte Carlo algorithm is employed for the Bayes estimation.  An intensive simulation study demonstrated the robustness of the minimum density power divergence and robust Bayes estimator over the conventional maximum likelihood estimator and Bayes estimator as the bias of robust estimators came out to be less than the conventional ones.  Further, robust testing of hypotheses is conducted by exploiting the Bayes factor, and the influence function is derived to assess the robustness of the estimators analytically.  Finally, a data analysis has been conducted to establish the utility of the theoretical results developed in this work.

This work can be extended to the non-parametric approach for inferential analysis.  The step-stress model can be reanalysed under a competing risk set-up.  The missing cause of failure analysis can also be conducted.  Efforts in this direction are in the pipeline, and we will report these findings soon.

\backmatter

\section*{Declarations}
\subsection*{Funding}
This research received no specific grant from funding agencies in the public, commercial, or not-for-profit sectors.
\subsection*{Conflict of interest}
The authors have no competing interests to declare that are relevant to the content of this article.
\subsection*{CRediT authorship contribution statement}
\noindent\textbf{Shanya Baghel:} Conceptualization, Formal analysis, Methodology, Software, Validation, Visualization, Writing - original draft.\\
\noindent \textbf{Shuvashree Mondal:}  Conceptualization, Methodology, Supervision, Validation, Visualization, Writing - review $\&$ editing.

\begin{appendices}

\section{Asymptotic distribution of MDE}\label{secA1}
Let $\bm{\theta}_0$ be true value of parameter 
$\bm{\theta}.$  With the help of the procedure followed by Calvino et al. \cite{calvino2021robustness}, the asymptotic distribution of MDE 
${\hat{\bm{\theta}} }_{\alpha}$ is given by
\begin{equation*}
  \sqrt{n}({\hat{\bm{\theta}} }_{\alpha} - {\bm{\theta}_0}) \xrightarrow[n\to \infty]{\mathscr{L}}N\Big(\bm{0}_{k+2}, J^{-1}_{\alpha}({\bm{\theta}_0}) K_{\alpha}({\bm{\theta}_0})J^{-1}_{\alpha}({\bm{\theta}_0}) \Big), 
\end{equation*}
where, 
\begin{align*}
J_{\alpha}({\bm{\theta}_0})&=u_s u^{T}_s p^{\gamma+1}_s+\sum_{i=1}^{k}\sum_{m=1}^{q_i}u_{im}u^{T}_{im}p^{\gamma+1}_{im}\\
K_{\alpha}({\bm{\theta}_0})&=u_s u^{T}_s p^{2\gamma+1}_s+\sum_{i=1}^{k}\sum_{m=1}^{q_i}u_{im}u^{T}_{im}p^{2\gamma+1}_{im}-\xi_{\alpha}(\bm{\theta}_0)\xi^{T}_{\alpha}(\bm{\theta}_0)\\
\xi_{\alpha}(\bm{\theta}_0)&=u_s p^{\gamma+1}_s+\sum_{i=1}^{k}\sum_{m=1}^{q_i}u_{im}p^{\gamma+1}_{im}\;;\; u=\frac{\partial}{\partial\bm{\theta}}ln\,p
\end{align*}

\section{Proof of Results}
\subsection{\textbf{Proof of Result \ref{res5}}}\label{ares5}
\begin{flalign*}
   \text{Denote,}\quad T^{(b)}_{\alpha}(F_{\bm{\theta}})&=\frac{\int\bm{\theta}\exp\big\{B_{\alpha}(\bm{\theta};F_{\bm{\theta}})\big\}\pi(\bm{\theta})\,d\bm{\theta}}{\int \exp\big\{B_{\alpha}(\bm{\theta};F_{\bm{\theta}})\big\}\pi(\bm{\theta})\,d\bm{\theta}}=\frac{T_1(F_{\bm{\theta}})}{T_2(F_{\bm{\theta}})}.\\
\text{Then,}\quad IF(t;T_{\alpha}^{(b)},F_{\bm{\theta}})&=\left.\frac{\partial}{\partial\epsilon}T_{\alpha}^{(b)}(U_{\epsilon})\right\vert_{\epsilon\to 0^{+}}\\
&=\left.\frac{T_2(U_{\epsilon})\frac{\partial}{\partial\epsilon}T_1(U_{\epsilon})-T_1(U_{\epsilon})\frac{\partial}{\partial\epsilon}T_2(U_{\epsilon})}{\left\{T_2(U_{\epsilon})\right\}^2}\right\vert_{\epsilon\to 0^{+}}\\
&=\frac{\int \bm{\theta}X_{\alpha}(\bm{\theta};t,f_{\bm{\theta}})\exp\left\{B_{\alpha}(\bm{\theta})\right\}\pi(\bm{\theta})d\bm{\theta}}{\int\exp\left\{B_{\alpha}(\bm{\theta})\right\}\pi(\bm{\theta})d\bm{\theta}} -\left[\frac{\int \bm{\theta}\exp\left\{ B_{\alpha}(\bm{\theta})\right\}\pi(\bm{\theta})d\bm{\theta}}{\int\exp\left\{B_{\alpha}(\bm{\theta})\right\}\pi(\bm{\theta})d\bm{\theta}}\right.\\
&\qquad\qquad\left.\times\frac{\int X_{\alpha}(\bm{\theta};t,f_{\bm{\theta}})\exp\left\{B_{\alpha}(\bm{\theta})\right\}\pi(\bm{\theta})d\bm{\theta}}{\int\exp\left\{B_{\alpha}(\bm{\theta})\right\}\pi(\bm{\theta})d\bm{\theta}}\right]\\
&=Cov_{(p)}\left(\bm{\theta},X_{\alpha}(\bm{\theta};t,f_{\bm{\theta}})\right),
&&
\end{flalign*}
\subsection{\textbf{Proof of Result \ref{res6}}}\label{ares6}
\begin{flalign*}
\text{Denote,}\quad	 T^{(\alpha)}_{\bm{\Theta}}(F_{\bm{\theta}_0})=\frac{\int_{\bm{\Theta}_0}\exp\big\{B_{\alpha}(\bm{\theta}\in\bm{\Theta}_0;F_{\bm{\theta}_0})\big\}\pi_0(\bm{\theta})\,d\bm{\theta}}{\int_{\bm{\Theta}_1}\exp\big\{B_{\alpha}(\bm{\theta}\in\bm{\Theta}_1;F_{\bm{\theta}_1})\big\}\pi_1(\bm{\theta})\,d\bm{\theta}}=\frac{T_0(\bm{\theta}\in\bm{\Theta}_0)}{T_1(\bm{\theta}\in\bm{\Theta}_1)}.
\end{flalign*}
\begin{flalign*}
\text{Then,}\; IF(t;T^{(\alpha)}_{\bm{\Theta}},F_{\bm{\theta}_0})&=\left.\frac{\partial(T^{(\alpha)}_{\bm{\Theta}}(U_{\epsilon}))}{\partial\epsilon}\right\vert_{\epsilon\to 0^{+}}.\\
&=\left[\frac{\int_{\bm{\Theta}_0}X_{\alpha}(\bm{\theta}\in\bm{\Theta}_0)\exp\big\{B_{\alpha}(\bm{\theta}\in\bm{\Theta}_0)\big\}\pi_0(\bm{\theta})\,d\bm{\theta}}{\int_{\bm{\Theta}_0}\exp\big\{B_{\alpha}(\bm{\theta}\in\bm{\Theta}_0)\big\}\pi_0(\bm{\theta})\,d\bm{\theta}}\times Y_{\alpha}(\bm{\Theta})\right]\\
&\qquad -\left[Y_{\alpha}(\bm{\Theta})\times\frac{\int_{\bm{\Theta}_1}X_{\alpha}(\bm{\theta}\in\bm{\Theta}_1)\exp\big\{B_{\alpha}(\bm{\theta}\in\bm{\Theta}_1)\big\}\pi_1(\bm{\theta})\,d\bm{\theta}}{\int_{\bm{\Theta}_1}\exp\big\{B_{\alpha}(\bm{\theta}\in\bm{\Theta}_1)\big\}\pi_1(\bm{\theta})\,d\bm{\theta}}\right]\\
&=Y_{\alpha}(\bm{\Theta})\bigg\{E\Big[X_{\alpha}(\bm{\theta}\in\bm{\Theta}_0)\Big]-E\Big[X_{\alpha}(\bm{\theta}\in\bm{\Theta}_1)\Big]\bigg\},\\
\text{where,}\;
X_{\alpha}(\bm{\theta}\in\bm{\Theta}_j)&=\left.\frac{\partial(B_{\alpha}(\bm{\theta}\in\bm{\Theta}_j;F_{\bm{\theta}_0}))}{\partial\epsilon}\right\vert_{\epsilon\to 0^{+}}\;;\;j=0,1.\\
	&=\frac{1}{\alpha}\left[\left\{\Delta^{(I_s)}_t-p_s(\bm{\theta}_0)\right\}p^{\alpha}_s(\bm{\theta}\in\bm{\Theta}_j)\right.\\
    &\qquad\left.+\sum_{i=1}^{k}\sum_{m=1}^{q_i}\left\{\Delta^{(I_{im})}_t-p_{im}(\bm{\theta}_0)\right\}p^{\alpha}_{im}(\bm{\theta}\in\bm{\Theta}_j)\right].
	&&
\end{flalign*}




\end{appendices}

\bibliography{sn-bibliography}% common bib file


\begin{thebibliography}{77}
\ifx \bisbn   \undefined \def \bisbn  #1{ISBN #1}\fi
\ifx \binits  \undefined \def \binits#1{#1}\fi
\ifx \bauthor  \undefined \def \bauthor#1{#1}\fi
\ifx \batitle  \undefined \def \batitle#1{#1}\fi
\ifx \bjtitle  \undefined \def \bjtitle#1{#1}\fi
\ifx \bvolume  \undefined \def \bvolume#1{\textbf{#1}}\fi
\ifx \byear  \undefined \def \byear#1{#1}\fi
\ifx \bissue  \undefined \def \bissue#1{#1}\fi
\ifx \bfpage  \undefined \def \bfpage#1{#1}\fi
\ifx \blpage  \undefined \def \blpage #1{#1}\fi
\ifx \burl  \undefined \def \burl#1{\textsf{#1}}\fi
\ifx \doiurl  \undefined \def \doiurl#1{\url{https://doi.org/#1}}\fi
\ifx \betal  \undefined \def \betal{\textit{et al.}}\fi
\ifx \binstitute  \undefined \def \binstitute#1{#1}\fi
\ifx \binstitutionaled  \undefined \def \binstitutionaled#1{#1}\fi
\ifx \bctitle  \undefined \def \bctitle#1{#1}\fi
\ifx \beditor  \undefined \def \beditor#1{#1}\fi
\ifx \bpublisher  \undefined \def \bpublisher#1{#1}\fi
\ifx \bbtitle  \undefined \def \bbtitle#1{#1}\fi
\ifx \bedition  \undefined \def \bedition#1{#1}\fi
\ifx \bseriesno  \undefined \def \bseriesno#1{#1}\fi
\ifx \blocation  \undefined \def \blocation#1{#1}\fi
\ifx \bsertitle  \undefined \def \bsertitle#1{#1}\fi
\ifx \bsnm \undefined \def \bsnm#1{#1}\fi
\ifx \bsuffix \undefined \def \bsuffix#1{#1}\fi
\ifx \bparticle \undefined \def \bparticle#1{#1}\fi
\ifx \barticle \undefined \def \barticle#1{#1}\fi
\bibcommenthead
\ifx \bconfdate \undefined \def \bconfdate #1{#1}\fi
\ifx \botherref \undefined \def \botherref #1{#1}\fi
\ifx \url \undefined \def \url#1{\textsf{#1}}\fi
\ifx \bchapter \undefined \def \bchapter#1{#1}\fi
\ifx \bbook \undefined \def \bbook#1{#1}\fi
\ifx \bcomment \undefined \def \bcomment#1{#1}\fi
\ifx \oauthor \undefined \def \oauthor#1{#1}\fi
\ifx \citeauthoryear \undefined \def \citeauthoryear#1{#1}\fi
\ifx \endbibitem  \undefined \def \endbibitem {}\fi
\ifx \bconflocation  \undefined \def \bconflocation#1{#1}\fi
\ifx \arxivurl  \undefined \def \arxivurl#1{\textsf{#1}}\fi
\csname PreBibitemsHook\endcsname

%%% 1
\bibitem[\protect\citeauthoryear{Prajapati
  et~al.}{2023}]{prajapati2023misspecification}
\begin{barticle}
\bauthor{\bsnm{Prajapati}, \binits{D.}},
\bauthor{\bsnm{Ling}, \binits{M.H.}},
\bauthor{\bsnm{Shing~Chan}, \binits{P.}},
\bauthor{\bsnm{Kundu}, \binits{D.}}:
\batitle{Misspecification of copula for one-shot devices under constant stress
  accelerated life-tests}.
\bjtitle{Proceedings of the Institution of Mechanical Engineers, Part O:
  Journal of Risk and Reliability}
\bvolume{237}(\bissue{4}),
\bfpage{725}--\blpage{740}
(\byear{2023})
\doiurl{10.1177/1748006X221108850}
\end{barticle}
\endbibitem

%%% 2
\bibitem[\protect\citeauthoryear{Fu et~al.}{2025}]{fu2025research}
\begin{botherref}
\oauthor{\bsnm{Fu}, \binits{X.}},
\oauthor{\bsnm{Wang}, \binits{B.}},
\oauthor{\bsnm{Su}, \binits{C.}},
\oauthor{\bsnm{Shao}, \binits{C.}}:
Research on the accelerated life test method for gaskets and the verification
  of the accuracy of the life prediction.
Journal of Pressure Vessel Technology,
1--16
(2025)
\doiurl{10.1115/1.4067678}
\end{botherref}
\endbibitem

%%% 3
\bibitem[\protect\citeauthoryear{Vali{\v{s}} et~al.}{2025}]{valivs2025light}
\begin{barticle}
\bauthor{\bsnm{Vali{\v{s}}}, \binits{D.}},
\bauthor{\bsnm{Forbelsk{\'a}}, \binits{M.}},
\bauthor{\bsnm{Vintr}, \binits{Z.}},
\bauthor{\bsnm{La}, \binits{Q.T.}},
\bauthor{\bsnm{Kohl}, \binits{Z.}}:
\batitle{Light emitting diode degradation and failure occurrence modelling
  based on accelerated life test}.
\bjtitle{Engineering Failure Analysis}
\bvolume{169},
\bfpage{109200}
(\byear{2025})
\doiurl{10.1016/j.engfailanal.2024.109200}
\end{barticle}
\endbibitem

%%% 4
\bibitem[\protect\citeauthoryear{Wang et~al.}{2025}]{wang2025review}
\begin{barticle}
\bauthor{\bsnm{Wang}, \binits{Y.}},
\bauthor{\bsnm{Chen}, \binits{X.}},
\bauthor{\bsnm{Zhang}, \binits{S.}},
\bauthor{\bsnm{Fan}, \binits{Z.}},
\bauthor{\bsnm{Hu}, \binits{J.}},
\bauthor{\bsnm{Yang}, \binits{C.}}:
\batitle{A review of modelling and data analysis methods for accelerated test}.
\bjtitle{Journal of Reliability Science and Engineering}
(\byear{2025})
\doiurl{10.1088/3050-2454/adb84e}
\end{barticle}
\endbibitem

%%% 5
\bibitem[\protect\citeauthoryear{Balakrishnan
  et~al.}{2022a}]{balakrishnan2022non}
\begin{barticle}
\bauthor{\bsnm{Balakrishnan}, \binits{N.}},
\bauthor{\bsnm{Jaenada}, \binits{M.}},
\bauthor{\bsnm{Pardo}, \binits{L.}}:
\batitle{Non-destructive one-shot device testing under step-stress model with
  weibull lifetime distributions}.
\bjtitle{arXiv preprint arXiv:2208.02674}
(\byear{2022})
\doiurl{10.48550/arXiv.2208.02674}
\end{barticle}
\endbibitem

%%% 6
\bibitem[\protect\citeauthoryear{Balakrishnan
  et~al.}{2022b}]{balakrishnan2022restricted}
\begin{barticle}
\bauthor{\bsnm{Balakrishnan}, \binits{N.}},
\bauthor{\bsnm{Jaenada}, \binits{M.}},
\bauthor{\bsnm{Pardo}, \binits{L.}}:
\batitle{The restricted minimum density power divergence estimator for
  non-destructive one-shot device testing the under step-stress model with
  exponential lifetimes}.
\bjtitle{arXiv preprint arXiv:2205.07103}
(\byear{2022})
\doiurl{10.48550/arXiv.2205.07103}
\end{barticle}
\endbibitem

%%% 7
\bibitem[\protect\citeauthoryear{Balakrishnan
  et~al.}{2023}]{balakrishnan2023robust}
\begin{barticle}
\bauthor{\bsnm{Balakrishnan}, \binits{N.}},
\bauthor{\bsnm{Castilla}, \binits{E.}},
\bauthor{\bsnm{Jaenada}, \binits{M.}},
\bauthor{\bsnm{Pardo}, \binits{L.}}:
\batitle{Robust inference for nondestructive one-shot device testing under
  step-stress model with exponential lifetimes}.
\bjtitle{Quality and Reliability Engineering International}
\bvolume{39}(\bissue{4}),
\bfpage{1192}--\blpage{1222}
(\byear{2023})
\doiurl{10.1002/qre.3287}
\end{barticle}
\endbibitem

%%% 8
\bibitem[\protect\citeauthoryear{Balakrishnan
  et~al.}{2024}]{balakrishnan2024non}
\begin{barticle}
\bauthor{\bsnm{Balakrishnan}, \binits{N.}},
\bauthor{\bsnm{Jaenada}, \binits{M.}},
\bauthor{\bsnm{Pardo}, \binits{L.}}:
\batitle{Non-destructive one-shot device test under step-stress experiment with
  lognormal lifetime distribution}.
\bjtitle{Journal of Computational and Applied Mathematics}
\bvolume{437},
\bfpage{115483}
(\byear{2024})
\doiurl{10.1016/j.cam.2023.115483}
\end{barticle}
\endbibitem

%%% 9
\bibitem[\protect\citeauthoryear{Ling}{2019}]{ling2019optimal}
\begin{barticle}
\bauthor{\bsnm{Ling}, \binits{M.H.}}:
\batitle{Optimal design of simple step-stress accelerated life tests for
  one-shot devices under exponential distributions}.
\bjtitle{Probability in the Engineering and Informational Sciences}
\bvolume{33}(\bissue{1}),
\bfpage{121}--\blpage{135}
(\byear{2019})
\doiurl{10.1017/S0269964818000049}
\end{barticle}
\endbibitem

%%% 10
\bibitem[\protect\citeauthoryear{Ling and Hu}{2020}]{ling2020optimal}
\begin{barticle}
\bauthor{\bsnm{Ling}, \binits{M.H.}},
\bauthor{\bsnm{Hu}, \binits{X.}}:
\batitle{Optimal design of simple step-stress accelerated life tests for
  one-shot devices under weibull distributions}.
\bjtitle{Reliability Engineering \& System Safety}
\bvolume{193},
\bfpage{106630}
(\byear{2020})
\doiurl{10.1016/j.ress.2019.106630}
\end{barticle}
\endbibitem

%%% 11
\bibitem[\protect\citeauthoryear{Balakrishnan
  et~al.}{2024}]{balakrishnan2024step}
\begin{barticle}
\bauthor{\bsnm{Balakrishnan}, \binits{N.}},
\bauthor{\bsnm{Jaenada}, \binits{M.}},
\bauthor{\bsnm{Pardo}, \binits{L.}}:
\batitle{Step-stress tests for interval-censored data under gamma lifetime
  distribution}.
\bjtitle{Quality Engineering}
\bvolume{36}(\bissue{1}),
\bfpage{3}--\blpage{20}
(\byear{2024})
\doiurl{10.1080/08982112.2023.2199826}
\end{barticle}
\endbibitem

%%% 12
\bibitem[\protect\citeauthoryear{{René Van Dorp} and
  Mazzuchi}{2004}]{RENEVANDORP200455}
\begin{barticle}
\bauthor{\bsnm{{René Van Dorp}}, \binits{J.}},
\bauthor{\bsnm{Mazzuchi}, \binits{T.A.}}:
\batitle{A general bayes exponential inference model for accelerated life
  testing}.
\bjtitle{Journal of Statistical Planning and Inference}
\bvolume{119}(\bissue{1}),
\bfpage{55}--\blpage{74}
(\byear{2004})
\doiurl{10.1016/S0378-3758(02)00411-1}
\end{barticle}
\endbibitem

%%% 13
\bibitem[\protect\citeauthoryear{Kannan et~al.}{2010}]{kannan2010survival}
\begin{botherref}
\oauthor{\bsnm{Kannan}, \binits{N.}},
\oauthor{\bsnm{Kundu}, \binits{D.}},
\oauthor{\bsnm{Balakrishnan}, \binits{N.}}:
Survival models for step-stress experiments with lagged effects.
Advances in Degradation Modeling: Applications to Reliability, Survival
  Analysis, and Finance,
355--369
(2010)
\doiurl{10.1007/978-0-8176-4924-1_23}
\end{botherref}
\endbibitem

%%% 14
\bibitem[\protect\citeauthoryear{Yao and Luo}{2013}]{yao2013step}
\begin{bchapter}
\bauthor{\bsnm{Yao}, \binits{J.-y.}},
\bauthor{\bsnm{Luo}, \binits{R.-m.}}:
\bctitle{Step-stress accelerated degradation test model of storage life based
  on lagged effect for electronic products}.
In: \bbtitle{The 19th International Conference on Industrial Engineering and
  Engineering Management: Engineering Management},
pp. \bfpage{541}--\blpage{550}
(\byear{2013}).
\doiurl{10.1007/978-3-642-38433-2-59} .
\bcomment{Springer}
\end{bchapter}
\endbibitem

%%% 15
\bibitem[\protect\citeauthoryear{Beltrami}{2017}]{beltrami2017weibull}
\begin{barticle}
\bauthor{\bsnm{Beltrami}, \binits{J.}}:
\batitle{Weibull lagged effect step-stress model with competing risks}.
\bjtitle{Communications in Statistics-Theory and Methods}
\bvolume{46}(\bissue{11}),
\bfpage{5419}--\blpage{5442}
(\byear{2017})
\doiurl{10.1080/03610926.2015.1102283}
\end{barticle}
\endbibitem

%%% 16
\bibitem[\protect\citeauthoryear{Qiao and Gui}{2023}]{qiao2023inference}
\begin{barticle}
\bauthor{\bsnm{Qiao}, \binits{Y.}},
\bauthor{\bsnm{Gui}, \binits{W.}}:
\batitle{Inference for cumulative risk model under step-stress experiments and
  its application in nanocrystalline data}.
\bjtitle{Proceedings of the Institution of Mechanical Engineers, Part O:
  Journal of Risk and Reliability}
\bvolume{237}(\bissue{1}),
\bfpage{195}--\blpage{209}
(\byear{2023})
\doiurl{10.1177/1748006X211072643}
\end{barticle}
\endbibitem

%%% 17
\bibitem[\protect\citeauthoryear{Mun et~al.}{2019}]{mun2019bayesian}
\begin{botherref}
\oauthor{\bsnm{Mun}, \binits{B.M.}},
\oauthor{\bsnm{Lee}, \binits{C.}},
\oauthor{\bsnm{Jang}, \binits{S.-g.}},
\oauthor{\bsnm{Ryu}, \binits{B.T.}},
\oauthor{\bsnm{Bae}, \binits{S.J.}}:
A bayesian approach for predicting functional reliability of one-shot devices.
International Journal of Industrial Engineering
\textbf{26}(1)
(2019)
\doiurl{10.23055/ijietap.2019.26.1.3638}
\end{botherref}
\endbibitem

%%% 18
\bibitem[\protect\citeauthoryear{Ariyo et~al.}{2022}]{ariyo2022bayesian}
\begin{barticle}
\bauthor{\bsnm{Ariyo}, \binits{O.}},
\bauthor{\bsnm{Lesaffre}, \binits{E.}},
\bauthor{\bsnm{Verbeke}, \binits{G.}},
\bauthor{\bsnm{Quintero}, \binits{A.}}:
\batitle{Bayesian model selection for longitudinal count data}.
\bjtitle{Sankhya B}
\bvolume{84}(\bissue{2}),
\bfpage{516}--\blpage{547}
(\byear{2022})
\doiurl{10.1007/s13571-021-00268-9}
\end{barticle}
\endbibitem

%%% 19
\bibitem[\protect\citeauthoryear{Abdel-Ghaly et~al.}{2023}]{abdel2023bayesian}
\begin{barticle}
\bauthor{\bsnm{Abdel-Ghaly}, \binits{A.}},
\bauthor{\bsnm{Aly}, \binits{H.}},
\bauthor{\bsnm{Abdel-Rahman}, \binits{E.}}:
\batitle{Bayesian inference under ramp stress accelerated life testing using
  stan}.
\bjtitle{Sankhya B}
\bvolume{85}(\bissue{1}),
\bfpage{132}--\blpage{174}
(\byear{2023})
\doiurl{10.1007/s13571-022-00300-6}
\end{barticle}
\endbibitem

%%% 20
\bibitem[\protect\citeauthoryear{Allotey and Harel}{2023}]{allotey2023bayesian}
\begin{barticle}
\bauthor{\bsnm{Allotey}, \binits{P.}},
\bauthor{\bsnm{Harel}, \binits{O.}}:
\batitle{Bayesian spatial modeling of incomplete data with application to hiv
  prevalence in ghana}.
\bjtitle{Sankhya B}
\bvolume{85}(\bissue{2}),
\bfpage{307}--\blpage{329}
(\byear{2023})
\doiurl{10.1007/s13571-023-00308-6}
\end{barticle}
\endbibitem

%%% 21
\bibitem[\protect\citeauthoryear{Sanju et~al.}{2024}]{sanju2024evaluating}
\begin{barticle}
\bauthor{\bsnm{Sanju}},
\bauthor{\bsnm{Kumar}, \binits{V.}},
\bauthor{\bsnm{Kumari}, \binits{P.}}:
\batitle{Evaluating the performance of bayesian approach for imputing missing
  data under different missingness mechanism}.
\bjtitle{Sankhya B}
\bvolume{86}(\bissue{2}),
\bfpage{713}--\blpage{723}
(\byear{2024})
\doiurl{10.1007/s13571-024-00344-w}
\end{barticle}
\endbibitem

%%% 22
\bibitem[\protect\citeauthoryear{Ling et~al.}{2024}]{ling2024efficient}
\begin{barticle}
\bauthor{\bsnm{Ling}, \binits{M.}},
\bauthor{\bsnm{Ng}, \binits{H.}},
\bauthor{\bsnm{Shang}, \binits{X.}},
\bauthor{\bsnm{Bae}, \binits{S.}}:
\batitle{Efficient bayesian inference for a defect rate based on completely
  censored data}.
\bjtitle{Applied Mathematical Modelling}
\bvolume{128},
\bfpage{123}--\blpage{136}
(\byear{2024})
\doiurl{10.1016/j.apm.2024.01.022}
\end{barticle}
\endbibitem

%%% 23
\bibitem[\protect\citeauthoryear{Salem and Salah}{2024}]{salem2024bayesian}
\begin{botherref}
\oauthor{\bsnm{Salem}, \binits{M.}},
\oauthor{\bsnm{Salah}, \binits{R.N.}}:
Bayesian inference for one-shot devices with weibull dependent failure modes
  using copulas.
Journal of the Indian Society for Probability and Statistics,
1--23
(2024)
\doiurl{10.1007/s41096-024-00222-8}
\end{botherref}
\endbibitem

%%% 24
\bibitem[\protect\citeauthoryear{Kumari and Sharma}{2024}]{kumari2024bayes}
\begin{botherref}
\oauthor{\bsnm{Kumari}, \binits{A.}},
\oauthor{\bsnm{Sharma}, \binits{V.K.}}:
Bayes estimation of defective proportion for single shot device testing data
  with information on masking and manufacturing defects.
Proceedings of the Institution of Mechanical Engineers, Part O: Journal of Risk
  and Reliability,
1748006--241299018
(2024)
\doiurl{10.1177/1748006X241299}
\end{botherref}
\endbibitem

%%% 25
\bibitem[\protect\citeauthoryear{Quigley et~al.}{2009}]{quigley2009empirical}
\begin{bchapter}
\bauthor{\bsnm{Quigley}, \binits{J.}},
\bauthor{\bsnm{Bedford}, \binits{T.}},
\bauthor{\bsnm{Walls}, \binits{L.}}:
\bctitle{Empirical bayes estimates of development reliability for one shot
  devices}.
In: \bbtitle{Safety and Reliability},
vol. \bseriesno{29},
pp. \bfpage{35}--\blpage{46}
(\byear{2009}).
\doiurl{10.1080/09617353.2009.11690888} .
\bcomment{Taylor \& Francis}
\end{bchapter}
\endbibitem

%%% 26
\bibitem[\protect\citeauthoryear{Lee et~al.}{2014}]{lee2014study}
\begin{barticle}
\bauthor{\bsnm{Lee}, \binits{Y.H.}},
\bauthor{\bsnm{Lee}, \binits{K.S.}},
\bauthor{\bsnm{Lee}, \binits{H.J.}},
\bauthor{\bsnm{Kim}, \binits{S.M.}},
\bauthor{\bsnm{Moon}, \binits{K.S.}}:
\batitle{A study of economical sample size for reliability test of one-shot
  device with bayesian techniques}.
\bjtitle{Journal of Applied Reliability}
\bvolume{14}(\bissue{3}),
\bfpage{162}--\blpage{168}
(\byear{2014})
\end{barticle}
\endbibitem

%%% 27
\bibitem[\protect\citeauthoryear{Balakrishnan
  et~al.}{2015}]{balakrishnan2015bayesian}
\begin{barticle}
\bauthor{\bsnm{Balakrishnan}, \binits{N.}},
\bauthor{\bsnm{So}, \binits{H.Y.}},
\bauthor{\bsnm{Ling}, \binits{M.H.}}:
\batitle{A bayesian approach for one-shot device testing with exponential
  lifetimes under competing risks}.
\bjtitle{IEEE Transactions on Reliability}
\bvolume{65}(\bissue{1}),
\bfpage{469}--\blpage{485}
(\byear{2015})
\doiurl{10.1109/TR.2015.2440235}
\end{barticle}
\endbibitem

%%% 28
\bibitem[\protect\citeauthoryear{Sharma and
  Upadhyay}{2018}]{sharma2018hierarchical}
\begin{barticle}
\bauthor{\bsnm{Sharma}, \binits{R.}},
\bauthor{\bsnm{Upadhyay}, \binits{S.}}:
\batitle{A hierarchical bayes analysis for one-shot device testing experiment
  under the assumption of exponentiality}.
\bjtitle{Communications in statistics-simulation and computation}
\bvolume{47}(\bissue{5}),
\bfpage{1297}--\blpage{1314}
(\byear{2018})
\doiurl{10.1080/03610918.2017.1310233}
\end{barticle}
\endbibitem

%%% 29
\bibitem[\protect\citeauthoryear{Sharma et~al.}{2021}]{sharma2021hierarchical}
\begin{barticle}
\bauthor{\bsnm{Sharma}, \binits{R.}},
\bauthor{\bsnm{Srivastava}, \binits{R.}},
\bauthor{\bsnm{Upadhyay}, \binits{S.K.}}:
\batitle{A hierarchical bayes analysis and comparison of ph weibull and ph
  exponential models for one-shot device testing experiment}.
\bjtitle{International Journal of Reliability, Quality and Safety Engineering}
\bvolume{28}(\bissue{05}),
\bfpage{2150036}
(\byear{2021})
\doiurl{10.1142/S0218539321500364}
\end{barticle}
\endbibitem

%%% 30
\bibitem[\protect\citeauthoryear{Ashkamini et~al.}{2023}]{ashkamini2023bayes}
\begin{botherref}
\oauthor{\bsnm{Ashkamini}},
\oauthor{\bsnm{Sharma}, \binits{R.}},
\oauthor{\bsnm{Upadhyay}, \binits{S.K.}}:
Bayes analysis of one-shot device testing data with correlated failure modes
  using copula models.
Communications in Statistics-Simulation and Computation,
1--20
(2023)
\doiurl{10.1080/03610918.2023.2288802}
\end{botherref}
\endbibitem

%%% 31
\bibitem[\protect\citeauthoryear{Rougier and
  Duncan}{2024}]{rougier2024bayesian}
\begin{barticle}
\bauthor{\bsnm{Rougier}, \binits{J.}},
\bauthor{\bsnm{Duncan}, \binits{A.}}:
\batitle{Bayesian modeling and inference for one-shot experiments}.
\bjtitle{Technometrics}
\bvolume{66}(\bissue{1}),
\bfpage{55}--\blpage{64}
(\byear{2024})
\doiurl{10.1080/00401706.2023.2224524}
\end{barticle}
\endbibitem

%%% 32
\bibitem[\protect\citeauthoryear{Salah and Salem}{2025}]{salah2025point}
\begin{botherref}
\oauthor{\bsnm{Salah}, \binits{R.N.}},
\oauthor{\bsnm{Salem}, \binits{M.}}:
Point and interval estimation for one-shot devices under weibull distribution
  with dependent failure modes using copulas.
Communications in Statistics-Simulation and Computation,
1--16
(2025)
\doiurl{10.1080/03610918.2025.2459868}
\end{botherref}
\endbibitem

%%% 33
\bibitem[\protect\citeauthoryear{Ghosh and Basu}{2016}]{g2016}
\begin{barticle}
\bauthor{\bsnm{Ghosh}, \binits{A.}},
\bauthor{\bsnm{Basu}, \binits{A.}}:
\batitle{Robust bayes estimation using the density power divergence}.
\bjtitle{Annals of the Institute of Statistical Mathematics}
\bvolume{68},
\bfpage{413}--\blpage{437}
(\byear{2016})
\doiurl{10.1007/s10463-014-0499-0}
\end{barticle}
\endbibitem

%%% 34
\bibitem[\protect\citeauthoryear{Basu et~al.}{1998}]{basu1998robust}
\begin{barticle}
\bauthor{\bsnm{Basu}, \binits{A.}},
\bauthor{\bsnm{Harris}, \binits{I.R.}},
\bauthor{\bsnm{Hjort}, \binits{N.L.}},
\bauthor{\bsnm{Jones}, \binits{M.}}:
\batitle{Robust and efficient estimation by minimising a density power
  divergence}.
\bjtitle{Biometrika}
\bvolume{85}(\bissue{3}),
\bfpage{549}--\blpage{559}
(\byear{1998})
\doiurl{10.1093/biomet/85.3.549}
\end{barticle}
\endbibitem

%%% 35
\bibitem[\protect\citeauthoryear{Patra et~al.}{2013}]{patra2013power}
\begin{barticle}
\bauthor{\bsnm{Patra}, \binits{S.}},
\bauthor{\bsnm{Maji}, \binits{A.}},
\bauthor{\bsnm{Basu}, \binits{A.}},
\bauthor{\bsnm{Pardo}, \binits{L.}}:
\batitle{The power divergence and the density power divergence families: the
  mathematical connection}.
\bjtitle{Sankhya B}
\bvolume{75},
\bfpage{16}--\blpage{28}
(\byear{2013})
\doiurl{10.1007/s13571-012-0050-3}
\end{barticle}
\endbibitem

%%% 36
\bibitem[\protect\citeauthoryear{Hazra and Ghosh}{2024}]{hazra2024robust}
\begin{barticle}
\bauthor{\bsnm{Hazra}, \binits{A.}},
\bauthor{\bsnm{Ghosh}, \binits{A.}}:
\batitle{Robust statistical modeling of monthly rainfall: The minimum density
  power divergence approach}.
\bjtitle{Sankhya B}
\bvolume{86}(\bissue{1}),
\bfpage{241}--\blpage{279}
(\byear{2024})
\doiurl{10.1007/s13571-024-00324-0}
\end{barticle}
\endbibitem

%%% 37
\bibitem[\protect\citeauthoryear{G{\'o}rny and Cramer}{2020}]{gorny2020exact}
\begin{barticle}
\bauthor{\bsnm{G{\'o}rny}, \binits{J.}},
\bauthor{\bsnm{Cramer}, \binits{E.}}:
\batitle{On exact inferential results for a simple step-stress model under a
  time constraint}.
\bjtitle{Sankhya B}
\bvolume{82}(\bissue{2}),
\bfpage{201}--\blpage{239}
(\byear{2020})
\doiurl{10.1007/s13571-019-00188-9}
\end{barticle}
\endbibitem

%%% 38
\bibitem[\protect\citeauthoryear{Bedbur and Seiche}{2022}]{bedbur2022testing}
\begin{botherref}
\oauthor{\bsnm{Bedbur}, \binits{S.}},
\oauthor{\bsnm{Seiche}, \binits{T.}}:
Testing the validity of a link function assumption in repeated type-ii censored
  general step-stress experiments.
Sankhya B,
1--24
(2022)
\doiurl{10.1007/s13571-021-00250-5}
\end{botherref}
\endbibitem

%%% 39
\bibitem[\protect\citeauthoryear{Fan et~al.}{2009}]{fan2009bayesian}
\begin{barticle}
\bauthor{\bsnm{Fan}, \binits{T.-H.}},
\bauthor{\bsnm{Balakrishnan}, \binits{N.}},
\bauthor{\bsnm{Chang}, \binits{C.-C.}}:
\batitle{The bayesian approach for highly reliable electro-explosive devices
  using one-shot device testing}.
\bjtitle{Journal of Statistical Computation and Simulation}
\bvolume{79}(\bissue{9}),
\bfpage{1143}--\blpage{1154}
(\byear{2009})
\doiurl{10.1080/00949650802142592}
\end{barticle}
\endbibitem

%%% 40
\bibitem[\protect\citeauthoryear{Samanta et~al.}{2018}]{samanta2018order}
\begin{barticle}
\bauthor{\bsnm{Samanta}, \binits{D.}},
\bauthor{\bsnm{Kundu}, \binits{D.}},
\bauthor{\bsnm{Ganguly}, \binits{A.}}:
\batitle{Order restricted bayesian analysis of a simple step stress model}.
\bjtitle{Sankhya B}
\bvolume{80},
\bfpage{195}--\blpage{221}
(\byear{2018})
\doiurl{10.1007/s13571-017-0139-9}
\end{barticle}
\endbibitem

%%% 41
\bibitem[\protect\citeauthoryear{Mondal and Kundu}{2019}]{mondal2019point}
\begin{barticle}
\bauthor{\bsnm{Mondal}, \binits{S.}},
\bauthor{\bsnm{Kundu}, \binits{D.}}:
\batitle{Point and interval estimation of weibull parameters based on joint
  progressively censored data}.
\bjtitle{Sankhya b}
\bvolume{81},
\bfpage{1}--\blpage{25}
(\byear{2019})
\doiurl{10.1007/s13571-017-0134-1}
\end{barticle}
\endbibitem

%%% 42
\bibitem[\protect\citeauthoryear{Mondal and Kundu}{2020}]{mondal2020bayesian}
\begin{barticle}
\bauthor{\bsnm{Mondal}, \binits{S.}},
\bauthor{\bsnm{Kundu}, \binits{D.}}:
\batitle{Bayesian inference for weibull distribution under the balanced joint
  type-ii progressive censoring scheme}.
\bjtitle{American Journal of Mathematical and Management Sciences}
\bvolume{39}(\bissue{1}),
\bfpage{56}--\blpage{74}
(\byear{2020})
\doiurl{10.1080/01966324.2019.1579124}
\end{barticle}
\endbibitem

%%% 43
\bibitem[\protect\citeauthoryear{Pal et~al.}{2021}]{pal2021bayesian}
\begin{barticle}
\bauthor{\bsnm{Pal}, \binits{A.}},
\bauthor{\bsnm{Mitra}, \binits{S.}},
\bauthor{\bsnm{Kundu}, \binits{D.}}:
\batitle{Bayesian order-restricted inference of a weibull multi-step
  step-stress model}.
\bjtitle{Journal of statistical theory and practice}
\bvolume{15},
\bfpage{1}--\blpage{33}
(\byear{2021})
\doiurl{10.1007/s42519-020-00164-x}
\end{barticle}
\endbibitem

%%% 44
\bibitem[\protect\citeauthoryear{Wiedner and Han}{2021}]{wiedner2021bayesian}
\begin{barticle}
\bauthor{\bsnm{Wiedner}, \binits{C.}},
\bauthor{\bsnm{Han}, \binits{D.}}:
\batitle{Bayesian inference for the simple step-stress accelerated life tests
  under order-restriction}.
\bjtitle{Procedia Manufacturing}
\bvolume{55},
\bfpage{147}--\blpage{153}
(\byear{2021})
\doiurl{10.1016/j.promfg.2021.10.021}
\end{barticle}
\endbibitem

%%% 45
\bibitem[\protect\citeauthoryear{Thach and
  Bris}{2019}]{thach2019reparameterized}
\begin{bchapter}
\bauthor{\bsnm{Thach}, \binits{T.T.}},
\bauthor{\bsnm{Bris}, \binits{R.}}:
\bctitle{Reparameterized weibull distribution: a bayes study using hamiltonian
  monte carlo}.
In: \bbtitle{Proceedings of the 29th European Safety and Reliability
  Conference. Singapore: Research Publishing},
pp. \bfpage{997}--\blpage{1004}
(\byear{2019}).
\doiurl{10.3850/978-981-11-2724-3-0494-cd}
\end{bchapter}
\endbibitem

%%% 46
\bibitem[\protect\citeauthoryear{Neal et~al.}{2011}]{neal2011mcmc}
\begin{barticle}
\bauthor{\bsnm{Neal}, \binits{R.M.}}, \betal:
\batitle{Mcmc using hamiltonian dynamics}.
\bjtitle{Handbook of markov chain monte carlo}
\bvolume{2}(\bissue{11}),
\bfpage{2}
(\byear{2011})
\end{barticle}
\endbibitem

%%% 47
\bibitem[\protect\citeauthoryear{Neal}{1996}]{neal2012bayesian}
\begin{bbook}
\bauthor{\bsnm{Neal}, \binits{R.M.}}:
\bbtitle{Bayesian Learning for Neural Networks}.
\bpublisher{Springer},
\blocation{New York}
(\byear{1996}).
\doiurl{10.1007/978-1-4612-0745-0}
\end{bbook}
\endbibitem

%%% 48
\bibitem[\protect\citeauthoryear{Thanh~Thach and
  Bri{\v{s}}}{2021}]{thanh2021additive}
\begin{barticle}
\bauthor{\bsnm{Thanh~Thach}, \binits{T.}},
\bauthor{\bsnm{Bri{\v{s}}}, \binits{R.}}:
\batitle{An additive chen-weibull distribution and its applications in
  reliability modeling}.
\bjtitle{Quality and Reliability Engineering International}
\bvolume{37}(\bissue{1}),
\bfpage{352}--\blpage{373}
(\byear{2021})
\doiurl{10.1002/qre.2740}
\end{barticle}
\endbibitem

%%% 49
\bibitem[\protect\citeauthoryear{Abba and Wang}{2023}]{abba2023new}
\begin{botherref}
\oauthor{\bsnm{Abba}, \binits{B.}},
\oauthor{\bsnm{Wang}, \binits{H.}}:
A new failure times model for one and two failure modes system: A bayesian
  study with hamiltonian monte carlo simulation.
Proceedings of the Institution of Mechanical Engineers, Part O: Journal of Risk
  and Reliability,
1748006--221146367
(2023)
\doiurl{10.1177/1748006X221146367}
\end{botherref}
\endbibitem

%%% 50
\bibitem[\protect\citeauthoryear{Monnahan et~al.}{2017}]{monnahan2017faster}
\begin{barticle}
\bauthor{\bsnm{Monnahan}, \binits{C.C.}},
\bauthor{\bsnm{Thorson}, \binits{J.T.}},
\bauthor{\bsnm{Branch}, \binits{T.A.}}:
\batitle{Faster estimation of bayesian models in ecology using hamiltonian
  monte carlo}.
\bjtitle{Methods in Ecology and Evolution}
\bvolume{8}(\bissue{3}),
\bfpage{339}--\blpage{348}
(\byear{2017})
\doiurl{10.1111/2041-210X.12681}
\end{barticle}
\endbibitem

%%% 51
\bibitem[\protect\citeauthoryear{Abba et~al.}{2024}]{abba2024robust}
\begin{botherref}
\oauthor{\bsnm{Abba}, \binits{B.}},
\oauthor{\bsnm{Wu}, \binits{J.}},
\oauthor{\bsnm{Muhammad}, \binits{M.}}:
A robust multi-risk model and its reliability relevance: A bayes study with
  hamiltonian monte carlo methodology.
Reliability Engineering \& System Safety,
110310
(2024)
\doiurl{10.1016/j.ress.2024.110310}
\end{botherref}
\endbibitem

%%% 52
\bibitem[\protect\citeauthoryear{Jeffreys}{1973}]{jeffreys1973scientific}
\begin{bbook}
\bauthor{\bsnm{Jeffreys}, \binits{H.}}:
\bbtitle{Scientific Inference}.
\bpublisher{Cambridge University Press},
\blocation{Cambridge}
(\byear{1973})
\end{bbook}
\endbibitem

%%% 53
\bibitem[\protect\citeauthoryear{Jeffreys}{1935}]{jeffreys1935some}
\begin{bchapter}
\bauthor{\bsnm{Jeffreys}, \binits{H.}}:
\bctitle{Some tests of significance, treated by the theory of probability}.
In: \bbtitle{Mathematical Proceedings of the Cambridge Philosophical Society},
vol. \bseriesno{31(2)},
pp. \bfpage{203}--\blpage{222}
(\byear{1935}).
\doiurl{10.1017/S030500410001330X} .
\bcomment{Cambridge University Press}
\end{bchapter}
\endbibitem

%%% 54
\bibitem[\protect\citeauthoryear{Jeffreys}{1998}]{jeffreys1998theory}
\begin{bbook}
\bauthor{\bsnm{Jeffreys}, \binits{H.}}:
\bbtitle{The Theory of Probability}.
\bpublisher{OuP Oxford},
\blocation{Great Clarendon Street}
(\byear{1998})
\end{bbook}
\endbibitem

%%% 55
\bibitem[\protect\citeauthoryear{Bobotas and Kateri}{2015}]{bobotas2015step}
\begin{barticle}
\bauthor{\bsnm{Bobotas}, \binits{P.}},
\bauthor{\bsnm{Kateri}, \binits{M.}}:
\batitle{The step-stress tampered failure rate model under interval
  monitoring}.
\bjtitle{Statistical Methodology}
\bvolume{27},
\bfpage{100}--\blpage{122}
(\byear{2015})
\doiurl{10.1016/j.stamet.2015.06.002}
\end{barticle}
\endbibitem

%%% 56
\bibitem[\protect\citeauthoryear{Kateri and Kamps}{2015}]{kateri2015inference}
\begin{barticle}
\bauthor{\bsnm{Kateri}, \binits{M.}},
\bauthor{\bsnm{Kamps}, \binits{U.}}:
\batitle{Inference in step-stress models based on failure rates}.
\bjtitle{Statistical Papers}
\bvolume{56},
\bfpage{639}--\blpage{660}
(\byear{2015})
\doiurl{10.1007/s00362-014-0601-y}
\end{barticle}
\endbibitem

%%% 57
\bibitem[\protect\citeauthoryear{Pal et~al.}{2021}]{pal2021simple}
\begin{botherref}
\oauthor{\bsnm{Pal}, \binits{A.}},
\oauthor{\bsnm{Samanta}, \binits{D.}},
\oauthor{\bsnm{Mitra}, \binits{S.}},
\oauthor{\bsnm{Kundu}, \binits{D.}}:
A simple step-stress model for lehmann family of distributions.
Advances in Statistics-Theory and Applications: Honoring the Contributions of
  Barry C. Arnold in Statistical Science,
315--343
(2021)
\doiurl{10.1007/978-3-030-62900-7_16}
\end{botherref}
\endbibitem

%%% 58
\bibitem[\protect\citeauthoryear{Kannan and Kundu}{2020}]{kannan2020simple}
\begin{barticle}
\bauthor{\bsnm{Kannan}, \binits{N.}},
\bauthor{\bsnm{Kundu}, \binits{D.}}:
\batitle{Simple step-stress models with a cure fraction}.
\bjtitle{Brazilian Journal of Probability and Statistics}
\bvolume{34}(\bissue{1}),
\bfpage{2}--\blpage{17}
(\byear{2020})
\doiurl{https://www.jstor.org/stable/26924216}
\end{barticle}
\endbibitem

%%% 59
\bibitem[\protect\citeauthoryear{Zhu}{2010}]{zhu2010optimal}
\begin{bbook}
\bauthor{\bsnm{Zhu}, \binits{Y.}}:
\bbtitle{Optimal Design and Equivalency of Accelerated Life Testing Plans}.
\bpublisher{Rutgers The State University of New Jersey, School of Graduate
  Studies},
\blocation{New Jersey}
(\byear{2010})
\end{bbook}
\endbibitem

%%% 60
\bibitem[\protect\citeauthoryear{Baghel and Mondal}{2024}]{baghel2024analysis}
\begin{barticle}
\bauthor{\bsnm{Baghel}, \binits{S.}},
\bauthor{\bsnm{Mondal}, \binits{S.}}:
\batitle{Analysis of one-shot device testing data under logistic-exponential
  lifetime distribution with an application to seer gallbladder cancer data}.
\bjtitle{Applied Mathematical Modelling}
\bvolume{126},
\bfpage{159}--\blpage{184}
(\byear{2024})
\doiurl{10.1016/j.apm.2023.10.037}
\end{barticle}
\endbibitem

%%% 61
\bibitem[\protect\citeauthoryear{Balakrishnan and
  Castilla}{2024}]{balakrishnan2024robust}
\begin{barticle}
\bauthor{\bsnm{Balakrishnan}, \binits{N.}},
\bauthor{\bsnm{Castilla}, \binits{E.}}:
\batitle{Robust inference for destructive one-shot device test data under
  weibull lifetimes and competing risks}.
\bjtitle{Journal of Computational and Applied Mathematics}
\bvolume{437},
\bfpage{115452}
(\byear{2024})
\doiurl{10.1016/j.cam.2023.115452}
\end{barticle}
\endbibitem

%%% 62
\bibitem[\protect\citeauthoryear{Baghel and Mondal}{2024}]{baghel2024robust}
\begin{barticle}
\bauthor{\bsnm{Baghel}, \binits{S.}},
\bauthor{\bsnm{Mondal}, \binits{S.}}:
\batitle{Robust estimation of dependent competing risk model under interval
  monitoring and determining optimal inspection intervals}.
\bjtitle{Applied Stochastic Models in Business and Industry}
(\byear{2024})
\doiurl{10.1002/asmb.2854}
\end{barticle}
\endbibitem

%%% 63
\bibitem[\protect\citeauthoryear{Lee and Morris~A.}{1985}]{le1985}
\begin{barticle}
\bauthor{\bsnm{Lee}, \binits{H.L.}},
\bauthor{\bsnm{Morris~A.}, \binits{C.}}:
\batitle{A multinomial logit model for the spatial distribution of hospital
  utilization}.
\bjtitle{Journal of Business \& Economic Statistics}
\bvolume{3}(\bissue{2}),
\bfpage{159}--\blpage{168}
(\byear{1985})
\doiurl{10.1080/07350015.1985.10509445}
\end{barticle}
\endbibitem

%%% 64
\bibitem[\protect\citeauthoryear{Thach and Bris}{2020}]{thach2020improved}
\begin{barticle}
\bauthor{\bsnm{Thach}, \binits{T.T.}},
\bauthor{\bsnm{Bris}, \binits{R.}}:
\batitle{Improved new modified weibull distribution: A bayes study using
  hamiltonian monte carlo simulation}.
\bjtitle{Proceedings of the Institution of Mechanical Engineers, Part O:
  Journal of Risk and Reliability}
\bvolume{234}(\bissue{3}),
\bfpage{496}--\blpage{511}
(\byear{2020})
\doiurl{10.1177/1748006X19896740}
\end{barticle}
\endbibitem

%%% 65
\bibitem[\protect\citeauthoryear{Thomas and Tu}{2021}]{thomas2021learning}
\begin{barticle}
\bauthor{\bsnm{Thomas}, \binits{S.}},
\bauthor{\bsnm{Tu}, \binits{W.}}:
\batitle{Learning hamiltonian monte carlo in r}.
\bjtitle{The American Statistician}
\bvolume{75}(\bissue{4}),
\bfpage{403}--\blpage{413}
(\byear{2021})
\doiurl{10.1080/00031305.2020.1865198}
\end{barticle}
\endbibitem

%%% 66
\bibitem[\protect\citeauthoryear{Ghosh et~al.}{2006}]{ghosh2006introduction}
\begin{bbook}
\bauthor{\bsnm{Ghosh}, \binits{J.K.}},
\bauthor{\bsnm{Delampady}, \binits{M.}},
\bauthor{\bsnm{Samanta}, \binits{T.}}:
\bbtitle{An Introduction to Bayesian Analysis: Theory and Methods}
vol. \bseriesno{725}.
\bpublisher{Springer},
\blocation{London}
(\byear{2006})
\end{bbook}
\endbibitem

%%% 67
\bibitem[\protect\citeauthoryear{Kass and Raftery}{1995}]{kass1995bayes}
\begin{barticle}
\bauthor{\bsnm{Kass}, \binits{R.E.}},
\bauthor{\bsnm{Raftery}, \binits{A.E.}}:
\batitle{Bayes factors}.
\bjtitle{Journal of the american statistical association}
\bvolume{90}(\bissue{430}),
\bfpage{773}--\blpage{795}
(\byear{1995})
\doiurl{10.1080/01621459.1995.10476572}
\end{barticle}
\endbibitem

%%% 68
\bibitem[\protect\citeauthoryear{Balakrishnan
  et~al.}{2019a}]{balakrishnan2019robust}
\begin{barticle}
\bauthor{\bsnm{Balakrishnan}, \binits{N.}},
\bauthor{\bsnm{Castilla}, \binits{E.}},
\bauthor{\bsnm{Mart{\'\i}n}, \binits{N.}},
\bauthor{\bsnm{Pardo}, \binits{L.}}:
\batitle{Robust estimators for one-shot device testing data under gamma
  lifetime model with an application to a tumor toxicological data}.
\bjtitle{Metrika}
\bvolume{82}(\bissue{8}),
\bfpage{991}--\blpage{1019}
(\byear{2019})
\doiurl{10.1007/s00184-019-00718-5}
\end{barticle}
\endbibitem

%%% 69
\bibitem[\protect\citeauthoryear{Balakrishnan
  et~al.}{2019b}]{balakrishnan2019robust1}
\begin{barticle}
\bauthor{\bsnm{Balakrishnan}, \binits{N.}},
\bauthor{\bsnm{Castilla}, \binits{E.}},
\bauthor{\bsnm{Mart{\'\i}n}, \binits{N.}},
\bauthor{\bsnm{Pardo}, \binits{L.}}:
\batitle{Robust inference for one-shot device testing data under weibull
  lifetime model}.
\bjtitle{IEEE transactions on Reliability}
\bvolume{69}(\bissue{3}),
\bfpage{937}--\blpage{953}
(\byear{2019})
\doiurl{10.1109/TR.2019.2954385}
\end{barticle}
\endbibitem

%%% 70
\bibitem[\protect\citeauthoryear{Balakrishnan
  et~al.}{2019c}]{balakrishnan2019robust2}
\begin{barticle}
\bauthor{\bsnm{Balakrishnan}, \binits{N.}},
\bauthor{\bsnm{Castilla}, \binits{E.}},
\bauthor{\bsnm{Mart{\'\i}n}, \binits{N.}},
\bauthor{\bsnm{Pardo}, \binits{L.}}:
\batitle{Robust estimators and test statistics for one-shot device testing
  under the exponential distribution}.
\bjtitle{IEEE Transactions on Information Theory}
\bvolume{65}(\bissue{5}),
\bfpage{3080}--\blpage{3096}
(\byear{2019})
\doiurl{10.1109/TIT.2019.2903244}
\end{barticle}
\endbibitem

%%% 71
\bibitem[\protect\citeauthoryear{Balakrishnan
  et~al.}{2020}]{balakrishnan2020robust}
\begin{barticle}
\bauthor{\bsnm{Balakrishnan}, \binits{N.}},
\bauthor{\bsnm{Castilla}, \binits{E.}},
\bauthor{\bsnm{Mart{\'\i}n}, \binits{N.}},
\bauthor{\bsnm{Pardo}, \binits{L.}}:
\batitle{Robust inference for one-shot device testing data under exponential
  lifetime model with multiple stresses}.
\bjtitle{Quality and Reliability Engineering International}
\bvolume{36}(\bissue{6}),
\bfpage{1916}--\blpage{1930}
(\byear{2020})
\doiurl{10.1002/qre.2665}
\end{barticle}
\endbibitem

%%% 72
\bibitem[\protect\citeauthoryear{Basak et~al.}{2021}]{bas2021}
\begin{barticle}
\bauthor{\bsnm{Basak}, \binits{S.}},
\bauthor{\bsnm{Basu}, \binits{A.}},
\bauthor{\bsnm{Jones}, \binits{M.}}:
\batitle{On the ‘optimal’density power divergence tuning parameter}.
\bjtitle{Journal of Applied Statistics}
\bvolume{48}(\bissue{3}),
\bfpage{536}--\blpage{556}
(\byear{2021})
\doiurl{10.1080/02664763.2020.1736524}
\end{barticle}
\endbibitem

%%% 73
\bibitem[\protect\citeauthoryear{Castilla and Chocano}{2022}]{cas2022}
\begin{botherref}
\oauthor{\bsnm{Castilla}, \binits{E.}},
\oauthor{\bsnm{Chocano}, \binits{P.J.}}:
On the choice of the optimal tuning parameter in robust one-shot device testing
  analysis.
Trends in Mathematical, Information and Data Sciences: A Tribute to Leandro
  Pardo,
169--180
(2022)
\doiurl{10.1007/978-3-031-04137-2_16}
\end{botherref}
\endbibitem

%%% 74
\bibitem[\protect\citeauthoryear{Sugasawa and
  Yonekura}{2021}]{sugasawa2021selection}
\begin{barticle}
\bauthor{\bsnm{Sugasawa}, \binits{S.}},
\bauthor{\bsnm{Yonekura}, \binits{S.}}:
\batitle{On selection criteria for the tuning parameter in robust divergence}.
\bjtitle{Entropy}
\bvolume{23}(\bissue{9}),
\bfpage{1147}
(\byear{2021})
\doiurl{10.3390/e23091147}
\end{barticle}
\endbibitem

%%% 75
\bibitem[\protect\citeauthoryear{Yonekura and
  Sugasawa}{2023}]{yonekura2023adaptation}
\begin{barticle}
\bauthor{\bsnm{Yonekura}, \binits{S.}},
\bauthor{\bsnm{Sugasawa}, \binits{S.}}:
\batitle{Adaptation of the tuning parameter in general bayesian inference with
  robust divergence}.
\bjtitle{Statistics and Computing}
\bvolume{33}(\bissue{2}),
\bfpage{39}
(\byear{2023})
\doiurl{10.1007/s11222-023-10205-7}
\end{barticle}
\endbibitem

%%% 76
\bibitem[\protect\citeauthoryear{Warwick and Jones}{2005}]{w2005}
\begin{barticle}
\bauthor{\bsnm{Warwick}, \binits{J.}},
\bauthor{\bsnm{Jones}, \binits{M.}}:
\batitle{Choosing a robustness tuning parameter}.
\bjtitle{Journal of Statistical Computation and Simulation}
\bvolume{75}(\bissue{7}),
\bfpage{581}--\blpage{588}
(\byear{2005})
\doiurl{10.1080/00949650412331299120}
\end{barticle}
\endbibitem

%%% 77
\bibitem[\protect\citeauthoryear{Calvino et~al.}{2021}]{calvino2021robustness}
\begin{barticle}
\bauthor{\bsnm{Calvino}, \binits{A.}},
\bauthor{\bsnm{Martin}, \binits{N.}},
\bauthor{\bsnm{Pardo}, \binits{L.}}:
\batitle{Robustness of minimum density power divergence estimators and
  wald-type test statistics in loglinear models with multinomial sampling}.
\bjtitle{Journal of Computational and Applied Mathematics}
\bvolume{386},
\bfpage{113214}
(\byear{2021})
\doiurl{10.1016/j.cam.2020.113214}
\end{barticle}
\endbibitem

\end{thebibliography}
\end{document}